\documentclass[preprint]{aastex}
\singlespace

\begin{document}

\title{Gas Distribution and Starburst Activity in the Widely Separated Interacting Galaxies NGC 6670}

\author{Wei-Hao Wang\altaffilmark{1,4}
\email{wang@ifa.hawaii.edu}
K. Y. Lo\altaffilmark{1}
\email{kyl@asiaa.sinica.edu.tw}
Yu Gao\altaffilmark{2}
\email{gao@ipac.caltech.edu}
and Robert A. Gruendl\altaffilmark{3}
\email{gruendl@astro.uiuc.edu}}
\altaffiltext{1}{Academia Sinica Institute of Astronomy and Astrophysics, P.O. Box 1-87,
Nankang, Taipei, Taiwan}
\altaffiltext{2}{IPAC, MS 100-22, Caltech, 770S Wilson Ave., Pasadena, CA 91125, USA}
\altaffiltext{3}{Laboratory for Astronomical Imaging, Department of Astronomy,
University of Illinois, 1002 W. Green Street, Urbana, IL 61801, USA}
\altaffiltext{4}{Current address: Institute for Astronomy, University of Hawaii, 2680 Woodlawn Drive,
Honolulu, HI 96822, USA}

\begin{abstract}
We present high resolution \ion{H}{1} 21 cm line, 20 cm radio continuum, and CO(1--0) line observations
of the luminous infrared galaxy NGC 6670.   NGC 6670 consists of two edge-on disk galaxies (NGC 6670E
and NGC 6670W) with a projected nuclear separation of $\sim 16$ kpc.   While there are no optically identified 
tidal features and the molecular disks are not distorted much, we have detected a 90 kpc long \ion{H}{1} tail 
which suggests that the galaxies are interacting and have already experienced at least one close encounter.
The galaxies appear to be gravitationally bound and in a prograde-prograde orbit, which is the most 
efficient for producing tidal tails.  Our observations suggest that the \ion{H}{1} at larger galactic radii has 
been ejected to form the tails and the remnant \ion{H}{1} disks have been perturbed by the interaction.  In
particular, we find that the \ion{H}{1} disk of NGC 6670E appears to have been nearly destroyed.
We conclude that the previous encounter between the galaxies had a large impact parameter and that the system 
is still in an early stage of interaction.

Even though NGC 6670 is in an early stage of interaction, we find that there is evidence for nuclear starbursts 
already present.  The CO(1--0) observations show large molecular gas reservoirs in the central regions and the 
20 cm radio continuum observations reveal enhanced star formation activity in the nuclear regions of both galaxies.  
The spatial extent of the 20 cm emission and the FIR-radio correlation further rule out active 
galactic nuclei as the source of the IR luminosity 
from NGC 6670.  We estimate the ratio $L_{\rm IR}/M_{\rm H_2}$, which is often used as an indicator of 
star formation efficiency, in the nuclear regions of NGC 6670E and NGC 6670W to be 18 and 11 
$L_{\sun}/M_{\sun}$, respectively.  The nuclear star formation efficiency of these two galaxies has been 
elevated to the level observed in other nearby starburst galaxies ($L_{\rm IR}/M_{\rm H_2}>10
L_{\sun}/M_{\sun}$).  Other indicators of starburst activity such as CO brightness temperature and infrared
surface brightness are also observed.  

\end{abstract}


\section{Introduction}
Luminous infrared galaxies (LIRGs) are galaxies with infrared luminosity
$L_{\rm IR}$(8--1000 $\rm \mu m$) greater than $10^{11} L_{\sun}$.  LIRGs are also the most
numerous sources with bolometric luminosity greater than $10^{11} L_{\sun}$ in the local universe
(see \citet{sanders96} for a recent review as well as the definition of infrared luminosity).   Observations
have shown that the large IR luminosity of LIRGs is primarily due to intense starbursts in the nuclear
regions and that dust-enshrouded active galactic nuclei (AGNs) may only play a minor role in some
ultraluminous (ULIRG, $L_{\rm IR}>10^{12} L_{\sun}$) systems 
\citep{joseph85,rieke85,condon91b,genzel98,downes98,smith98}.  For starbursts to produce
the infrared luminosity of LIRGs, more than $10^9 M_{\sun}$ of young massive stars are needed to
heat the interstellar dust.  This requires a concentration of at least $10^9 M_{\sun}$ of molecular gas
in a very small region ($\sim$~1 kpc) to form these stars (cf. \citet{lo87}).

Most LIRGs are found in interacting/merging systems \citep{sanders88,melnick90,murphy96,duc98}
that are rich in molecular gas \citep{sanders91,solomon92,solomon97,gao99}.  The fraction of the
LIRGs in interacting/merging systems increases with infrared luminosity, from $\sim 10\%$ at
$L_{\rm{IR}}=10^{10.5}-10^{11} L_{\sun}$ to $\sim 100\%$ at $L_{\rm{IR}}>10^{12} L_{\sun}$
\citep{sanders88,sanders96,melnick90,murphy96}.  This suggests that galaxy-galaxy interactions play
a major role in triggering starbursts in molecular gas-rich galaxies.  A commonly discussed interpretation
of the triggering mechanism is that, during the interaction, the gas clouds experience tidal torque produced
by either the host galaxy or the companion.  The clouds which lose their angular momentum flow into the
circum-nuclear regions of the host galaxies.  As a result, the concentration and compression of molecular
gas in the circum-nuclear regions give rise to starbursts.  Furthermore, during the final stage of the galaxies'
merging, clouds with small angular momentum may flow into the galactic nuclei and then trigger an AGN
or an ultraluminous starburst.  This scenario, however, has not been confirmed by observations.  Moreover, 
only a small fraction of optically selected interacting/merging systems are truly IR luminous or show other 
elevated tracers of star formation \citep{arp66,kennicutt87,bushouse88}.  There also exist LIRGs which are
gas-rich interacting galaxies but are not undergoing starbursts (e.g., Arp 302 \citep{lo97}).  Apparently, 
galactic interaction is not the only causal factor in the enhanced IR luminosity.

Recent numerical simulations of the merging of gas-rich spiral galaxies have shown how the gas and stars
may be redistributed during the interaction \citep{barnes91,mihos93,mihos96}.  The simulations can
predict a sharp rise and fall in  star formation rate (SFR) in the nuclear regions of galaxies, by assuming
the SFR is proportional to some power of the gas density (e.g., the Schmidt law).  On the other hand,
the physical meaning of the Schmidt law itself is still unclear and a range of power indices for the Schmidt
law are suggested by observations \citep{scoville86,kennicutt98,taniguchi98}.  It is difficult to assess whether
or not these simplified recipes which relate gas density to star formation produce realistic results.
Furthermore, \citet{jog92} showed that compression of giant molecular clouds due to the over-pressure
produced by \ion{H}{1} cloud-cloud collisions in colliding galaxies can produce starbursts in off-nuclear
regions.   On the other hand, simulations \citep{mihos93} fail to reproduce the gas concentrations in the
off-nuclear regions in systems like Arp 244 and Arp 299 (see, e.g., \citet{stanford90,sargent91,mirabel98}).
Therefore, a detailed understanding of where and when starbursts take place, as well as the physical
mechanism responsible for triggering them, remains unknown.

Previous observational studies of starbursts in LIRGs have mostly concentrated on ultraluminous or
advanced merging systems (e.g., \citet{lo87,sanders88,scoville91,genzel98}).  In such systems, the interstellar
medium (ISM) has already been highly disrupted by the tidal interaction and the starbursts.  Therefore, these
observations can only provide limited comparison with the models.  In order to trace the distribution of gas and
the evolution of starbursts along the merging process, we started a program of multi-wavelength observations to
study a sample of LIRGs in early/intermediate merging phases.  Partial results of this program can be found in
\citet{lo97}, \citet{gaoetal97}, \citet{hwang99}, \citet{gaoetal99}, \citet{lo2000}, and \citet{gao01}.  The goal is 
to sample the evolution of the ISM properties in LIRGs and to verify observationally the proposed triggering 
mechanisms of starbursts.

In this paper, we present interferometric observations of the \ion{H}{1} 21 cm line, the 20 cm radio
continuum, and the CO(1--0) line emission of NGC 6670 (UGC 11284).  NGC 6670 consists of two edge-on
disk galaxies with an angular separation of $\sim 27\arcsec$ between the two nuclei.  At a redshift (cz) of
8684 km $\rm s^{-1}$, its distance is 120 Mpc ($H_0$ = 75 km $\rm s^{-1}$ $\rm Mpc^{-1}$,
$q_0=0.5$) and the projected nuclear separation is $\sim 16$ kpc.  Near infrared (NIR) images of NGC 6670
show a slightly warped disk (the western galaxy, NGC 6670W) and another disk (the eastern one, 
NGC 6670E) with a mildly distorted shape.   The disks appear to just touch each other in the sky plane.
Because of the edge-on geometry and the lack of optical
tidal features, it is difficult to determine the interaction history, the orbital geometry of this system, or even
whether the galaxies are spatially related or superposed by chance.   As the neutral hydrogen gas has a more
extended distribution than stars and is less bound to the parent galaxies, it is a sensitive tracer of gravitational
perturbation.  Imaging the \ion{H}{1} emission provides the unique opportunity to verify the interaction 
between the galaxies of NGC 6670.

The infrared luminosity of NGC 6670 is $3.8\times10^{11} L_{\sun}$.  Single-dish measurements
show a total CO flux of 220 Jy km $\rm s^{-1}$, which corresponds to a molecular gas mass of
$\sim3.9\times10^{10} M_{\sun}$ \citep{gao96,gao99}.  The global ratio of $L_{\rm IR}/M_{\rm H_2}\equiv
\rm SFE$ is often used as an indication of star formation efficiency, the star formation rate per
unit molecular gas mass.  In NGC 6670, the global SFE is 10 $L_{\sun}/M_{\sun}$, only a factor of 2 higher
than the average SFE of the giant molecular clouds (GMCs) in the Milky Way's disk \citep{scoville89} and of
the isolated galaxies \citep{young91}.   Without knowing the distributions of molecular gas and star formation
activity, one cannot tell whether the large IR luminosity of NGC 6670 is due to nuclear starbursts 
($\rm SFE > 20$ $L_{\sun}/M_{\sun}$, \citep{scoville91,bryant99}) induced by the interaction, or simply 
due to the large reservoir of molecular gas with a rather ordinary SFE as in Arp 302 \citep{lo97}.  The 
Infrared Space Observatory (ISO) mid-infrared observations of NGC 6670 suggest that most of the 
mid-infrared emission comes from extended circum-nuclear starbursts but not from dust enshrouded AGNs
\citep{hwang99}.  Our CO(1--0) and 20 cm interferometric observations provide a higher resolution view that 
allows us to investigate the distribution of molecular gas and star formation.  

\section{Observation and Data Reduction}

\subsection{Optical and Near-Infrared Observations}
Broad-band optical $R$-band and near-infrared $J$-band images of the NGC~6670 system were acquired
using the 1-m telescope at Mt. Laguna Observatory.  $R$-band observations of NGC 6670 were taken
with a 2048$\times$2048 CCD on 1997 August 2.  These images were bias and flat-field corrected,
cosmic-ray hits were rejected, and the individual exposures were coadded to form the final image
with 720 seconds total integration time and $2\farcs4$ spatial resolution.

The $J$-band observations were obtained with the NIRIM camera \citep{meixner99} on
the 1-m telescope at Mt. Laguna Observatory on 1998 May 19.  The camera was configured
such that the $256\times256$ NICMOS 3 array was operating at a $1\arcsec$ per pixel plate
scale.  The objectives of these observations were to image the region around and between the
NGC 6670 system and the companion galaxy CGCG 301-032 (IRASF 18335+5949) roughly
$5\arcmin$ to the southeast.  To this end, a field $13\farcm5 \times 6\farcm8$ was imaged in
$94\times60$ second exposures.  The individual exposures were taken in a sequence that included 2
sky (off-source) exposures after every 3 on-source exposures.  A first-order sky subtraction
was achieved by subtracting the median of the four off-source exposures which temporally
bracketed each on-source exposure.  The spatial offsets of the resulting 53 on-source frames
were determined from field stars and then the zero-point offset between all overlapping frames
were calculated.  A non-linear least-squares solution was used to determine the best zero-point
offset for each frame \citep{regan95} before the frames were combined into a final mosaic
image of the entire field.  The typical resolution measured from stars in the final mosaic image was
$2\farcs7$

Additional optical broad-band images were acquired from the Issac Newton Group Archive.
These observations were made on 1995 May 24 with the 1-m Jacobus Kapteyn Telescope at
La Palma at $B$, $V$, and $I$-band, with integration times of 900, 600, and 900 seconds respectively.
These images, along with the associated bias and flat-field measurements, were retrieved from
the archive.  The images were bias and flat-field corrected and cosmic-ray hits were removed.
The resulting $B$, $V$, and $I$-band images have spatial resolutions of $1\farcs5$, $1\farcs3$, and
$1\farcs2$ respectively.

In addition, a near-infrared $K^{\prime}$-band image (D.-C. Kim and D. B. Sanders, private communication)
with $\sim 1\farcm5$ field of view, obtained with the University of Hawaii 88-inch telescope on Mauna 
Kea, is included for comparison with our radio images.  NGC 6670 was one of the targets in their 
optical/near-infrared imaging survey of extended Bright Galaxy Sample galaxies (see \citet{sanders95}; 
Sanders et al., in preparation).  Astrometric solutions for all the optical and near-IR observations were obtained 
by using stars from the Hubble Guide Star Catalog 1.2, the typical uncertainty in the registration of any of
the optical images is less than $0\farcs5$.

\subsection{21 cm \ion{H}{1} Observation}
\ion{H}{1} observations were obtained with the NRAO Very Large Array (VLA) C configuration on 1997
September 13.  The total usable integration time on NGC 6670 was nearly 3 hours.  The radio source 
3C 286 was observed for flux and bandpass calibration at the beginning of the observation.  The phase 
calibrator 3C 343 was observed for 60--90 seconds before and after every 40 minutes of on-source integration 
of NGC 6670.  The observing frequency was 1.38056 GHz, corresponding to cz=8650 km s$^{-1}$.
With on-line Hanning smoothing and a total of 63 channels over a bandwidth of 6.25 MHz, the channel
resolution was 21.8 km $\rm s^{-1}$.

The interferometric data were reduced with the NRAO \it AIPS \rm package.  The calibrated \it u-v \rm data
were inverted to image cubes using uniform weighting and Briggs' Robust Weighting \citep{briggs95}.  When
applying Robust Weighting, we chose the parameter \it ROBUST \rm = $+1.2$ for the imaging task \it IMAGR.
\rm The fitted synthesized beams are $16\farcs3\times11\farcs4$ at $51\fdg2$ position angle for
uniform weighting and $19\farcs1\times15\farcs6$ at $61\fdg5$ for Robust Weighting.  After the
\it CLEAN \rm algorithm was applied, an $11\farcs4$ circular restoring beam was used for the uniformly
weighted cube and an $18\farcs0 \times15\farcs6$ beam at $61\fdg5$ was used for the cube of Robust
Weighting.  A continuum map was obtained from each cube by averaging the line-free channels 5--18,
which corresponds to the heliocentric velocity between 9239 km $\rm s^{-1}$ and 8955 km $\rm s^{-1}$.
The continuum map was then subtracted from each cube to obtain a cube containing only \ion{H}{1} emission.
In the resulting continuum-free cubes of uniform and Robust Weighting, the noise levels ($\sigma$) are 0.50 
and 0.35 mJy $\rm beam^{-1}$ ($\sim 2.3$ K and 0.74 K brightness temperature), respectively.  A correction 
for the primary beam pattern was then applied to the continuum-free cubes.  Moment maps were then made 
from both cubes using the standard \it MOMNT \rm task to produce the integrated intensity maps, the 
intensity-weighted radial velocity maps, and the intensity-weighted velocity dispersion maps.

\subsection{20 cm Radio Continuum Observation}
NGC 6670 was observed at 20 cm with the B array configuration of the VLA on 1997 March 9.  Two 
intermediate frequencies were centered at 1.4649 and 1.3851 GHz with 50 MHz bandwidths.  The radio 
source 3C48 (0137$+$331) was observed for 6 minutes for flux and bandpass calibration.  The phase 
calibrator, 1927$+$612, was observed for 1.5 minutes before and after every 15 minutes of integration on 
NGC 6670.  The total integration time on NGC 6670 was 1 hour.  The NRAO $AIPS$ package was used 
for data reduction.  The visibilities were uniformly weighted, Fourier transformed, and then cleaned.  The 
``dirty'' beam was fitted to be $5\farcs9\times3\farcs1$ at an $83\fdg3$ position angle and a circular 
restoring beam of $3\farcs1$ was used after the \it CLEAN \rm algorithm had been applied.  The noise 
level in the restored map is $4.7\times10^{-2}$ mJy $\rm beam^{-1}$ which corresponds to 2.7 K 
brightness temperature.

\subsection{CO(1--0) Observations}
We observed NGC 6670 in the CO($J=1$--0) transition with the Berkeley-Illinois-Maryland Association
(BIMA) millimeter array \citep{welch96} in 1996 and 1997 using the 9-element C configuration and the
10-element B configuration.  The digital correlator was used with the widest available bandwidth to cover 
a velocity range of 1,600 km $\rm s^{-1}$, at a resolution of 8.4 km $\rm s^{-1}$.  The $\sim 2\arcmin$ 
primary beam was centered between the two galaxies to best cover the emission from this system which 
extends slightly more than $1\arcmin$. The observing frequency was 112.0567 GHz, corresponding to the 
CO(1--0) line at the median redshift cz=8,600 km $\rm s^{-1}$ of this galaxy system.  The nearby quasar,
1824$+$568, was observed for 8 minutes every half an hour to obtain the antenna phase and amplitude
gain calibration.  The strong quasar 3C454.3, and a planet (Mars or Uranus) were also observed for each
track to obtain a flux calibration.  Observations were obtained on seven different dates with typical system
temperatures between 250 and 800 K.  The total on-source integration time for NGC 6670 was 24 hours.

The interferometric data were reduced with the \it MIRIAD \rm data reduction package \citep{sault95}.
Data cubes of the CO brightness distribution, ${\rm T_b(x,y,V)}$, where x and y are the spatial coordinates
and V is the radial velocity were made at 20 km $\rm s^{-1}$ resolutions.  Two different spatial resolutions
were obtained by applying different weight functions to the visibilities prior to the Fourier transformation.  
First, all visibilities were weighted by the system temperature at the time of the observation, then two different
weight functions were applied to achieve two different resolution data cubes.  The first case was chosen to
produce a synthesized beam with roughly natural weighting but sidelobes in the dirty synthesized beam were
suppressed by using the Briggs' Robust Weighting \citep{briggs95} to determine the specific weighting of each
visibility.  In the second case Briggs' Robust Weighting was also used but an additional Gaussian taper was
applied to the visibilities to down-weight the long-baseline B-array data.

After the Fourier transform of the visibilities produced a dirty map, the \it MIRIAD CLEAN \rm algorithm was
used to remove sidelobe response and produce final CO cubes.  The final CO cubes cover a range from 
$-$500 to $+$500 km $\rm s^{-1}$ about the systemic velocity (cz=8,600 km $\rm s^{-1}$).  The cubes 
resulting from this reduction with the two different weighting schemes have final synthesized beams of 
$2\farcs7\times2\farcs1$ and $5\farcs4\times4\farcs6$.  The root-mean-square (rms) noise level is 
13 mJy $\rm beam^{-1}$ (0.2 K brightness temperature) in each channel of the high resolution clean cube 
and is 15 mJy $\rm beam^{-1}$ ( 0.05 K) in each channel of the low resolution clean cube.  Moment maps 
were then made to produce the integrated intensity maps and the intensity-weighted radial velocity maps.

\section{Results}

\subsection{Optical to Near-IR Morphology of NGC~6670}\label{optresults}
Based on previous observations, \citet{zwicky71} described the morphologic components of the NGC 6670
system as a ``post-eruptive quadruple system of three brush-like and one spherical compact'' galaxies.
Figure~\ref{opticals} shows optical and near-infrared ($B, V, R, I, J, K^{\prime}$-band) images of the NGC
6670 system.  These images reveal that the two largest galaxies in this system are a pair of nearly edge-on
spiral galaxies.   The relatively weak bulges of these two galaxies, compared to the disks, suggest that
these two galaxies are late type spiral galaxies.  To the southeast of the nucleus of NGC 6670E (the eastern 
spiral) there is a bright but fuzzy feature in shorter wavelengths.  At NIR wavelengths, especially in the
$K^{\prime}$-band, it appears more like a part of NGC 6670E's disk.  The different appearance between 
longer and shorter wavelengths in this region can be due to either dust extinction or a large amount of blue 
stars.  This feature could be a dusty region, or a young stellar tidal feature in the outer disk of 
NGC 6670E, or a separate dwarf galaxy.   Furthermore, the CO morphology, CO kinematics (\S~\ref{coresult}), and
radio continuum morphology (\S~\ref{20cmresult}) all suggest only two edge-on disks in this system.  It is most 
likely that the fuzzy feature in the southeast end of NGC 6670E is part of its disk or a young stellar tidal feature.

The western spiral galaxy (NGC 6670W) appears to be the more edge-on of the two spiral galaxies.
At NIR wavelengths a prominent bulge is evident, while at blue wavelengths the bulge and eastern portion 
of the disk are barely detected.  This is likely due to dust extinction in the eastern part of the disk.  We note 
that the heavily obscured region in the eastern part of the disk of NGC 6670W is also the location of the 
huge \ion{H}{1} concentration (the ``C concentration,'' see \S~\ref{h1}).   It is possible that this \ion{H}{1}
concentration is responsible for the extinction.  The western half of the disk is in general brighter than the 
eastern half especially at shorter wavelengths.  This may be an indication of recent star formation.    Both 
the eastern and western halves of the disk in NGC 6670W show significant curvature.  This morphology 
could be attributed to: 1) a warped disk, 2) a two-armed grand-design spiral with large pitch angle, inclined 
at nearly perfect edge-on, or 3) a tidal perturbation that is nearly in the plane of the galaxy.

In contrast to NGC 6670W, the nuclear region of NGC 6670E is clearly detected in all of the broad-band
images.  Furthermore, the disk appears less edge-on, with surface brightness variations that differ from one
band to the next.  These morphological variations could be attributed to dust extinction and/or massive star
formation.  On the other hand, if the emission detected at $B$-band traces star formation in the absence of
dust, then the star formation in the nuclear region of NGC 6670E is prolific in comparison to the 
nucleus/bulge of NGC 6670W.

The ``brush-like'' features noted by \citet{zwicky71} can all be attributed to the disks of the two spiral galaxies.  
One of these, the eastern portion of NGC 6670W disk, appears to extend to ``touch'' NGC
6670E.  This is most clear in the $K^{\prime}$-band image.  In all the broad-band images, the disks of both
NGC 6670E and NGC 6670W are enveloped in a faint ``halo'' of star light.  This ``common envelope,'' the
disturbed morphologies of both disks, and the apparent proximity of the disks suggest that the true separation
between the two galaxies is not much larger than the projected separation and the galaxies may be interacting.
On the other hand, none of these clues, along or together, prove that the two galaxies are interacting, much less 
in physical contact.

\subsection{\ion{H}{1} Results}
\subsubsection{Distribution of \ion{H}{1} in NGC 6670}\label{h1}
In Figure~\ref{h1moms}, we present the low resolution (robust weighting) \ion{H}{1} integrated intensity
map and the intensity-weighted velocity map.    Figure~\ref{h1unifmom0} shows an enlarged view of the
high resolution (uniform weighting) \ion{H}{1} integrated intensity.  For comparison, CO(1--0) integrated
intensity is also plotted in Figure~\ref{h1unifmom0} (see \S~\ref{codist}).  Several observed quantities are
summarized in Table~\ref{tab1}.

A previously unknown long \ion{H}{1} tail is revealed to the west of the stellar disks of NGC 6670
(hereafter named the ``W tail,''  see Fig.~\ref{h1moms}).  The W tail has a projected extension of  $\sim90$
kpc from the center between the two galaxies.  Furthermore, a smaller \ion{H}{1} tail extends $\sim25$
kpc away from NGC 6670 in the northeast direction (the ``NE tail'').   Neither tail has any obvious optical 
counterparts.  The existence of \ion{H}{1} tidal tails manifests that the galaxies in NGC 6670 are interacting,
and are not by chance superposed.

The \ion{H}{1} radial velocity map (Fig.~\ref{h1moms}b) and the \ion{H}{1} spectra (Fig.~\ref{h1spect}) 
show that in NGC 6670, the largest blueshift (8360 km $\rm s^{-1}$) appears in the W tail and the largest 
redshift (8820 km $\rm s^{-1}$) appears in the NE tail.  The velocity of the whole system increases systemically 
from the W tail to the NE tail.  The \ion{H}{1} velocity dispersion (cf. Fig.~\ref{h1spect} and Table~\ref{tab1}) 
in each tail is relatively small, between 11 km $\rm s^{-1}$ and 17 km $\rm s^{-1}$ .  The 
velocity of the W tail varies between 8432 and 8366 km $\rm s^{-1}$ (Fig.~\ref{h1spect}) and the velocity
difference between the end and the middle of the W tail is less than 20 km $\rm s^{-1}$.  The NE tail only
appears in the channels of 8825 and 8803 km $\rm s^{-1}$.  The opposite sense of velocity
and the relatively narrow velocity range of the \ion{H}{1} tails in NGC 6670 are very similar to the
\ion{H}{1} properties on the tails of the intermediate-stage merger Arp 244 ($The$ $Antennae$, \citet{hulst79}) 
and the early-stage merger NGC 4676 ($The$ $Mice$, \citet{hibbard96}).  On the other hand, only 25\%
of NGC 6670's \ion{H}{1} emission comes from the two tails, in contrast with Arp 244 and NGC 4676
which have more than 60\% of their \ion{H}{1} distributed along the tails.  With the trend found by
\citet{hibbard96} that late stage mergers have more \ion{H}{1} outside the stellar disks than early stage
mergers, NGC 6670 is most likely in an early stage of interaction.  This is consistent with the mild distortion
of the disks revealed at optical and NIR wavelengths.

The total \ion{H}{1} flux of NGC 6670 (including the tails), obtained by adding all the flux in the low resolution
channel maps (Fig.~\ref{h1chmap}), is 4.83 Jy km $\rm s^{-1}$.  
The single dish \ion{H}{1} flux of NGC 6670 measured by
\citet{martin91} is 4.19 Jy km $\rm s^{-1}$, which is comparable with our results.  Their result may be a lower
limit because their half-power beam width in the east-west direction ($4\arcmin$) is similar to the source extent
and they might have missed parts of the extended \ion{H}{1} tails.  In addition, the most recent results of
\citet{driel01} suggest that the \ion{H}{1} flux in Martin et al. may be systematically lower by up to 30\%.
Taking this into consideration, we tend to conclude that our VLA observation has probably detected almost 
all of  the \ion{H}{1} flux in this system.

The \ion{H}{1} intensity map (Fig.~\ref{h1moms}) shows that most of the \ion{H}{1} in NGC 6670 is
around the two separated stellar disks and forms a common \ion{H}{1} halo.  Three concentrations of
\ion{H}{1} gas can be identified near the stellar disks: the ``W concentration'' in western part of the NGC
6670W disk, the ``C concentration'' between the two galaxies, and the weaker ``E concentration'' in eastern
part of the NGC 6670E disk (Fig.~\ref{h1moms}a and Fig.~\ref{h1unifmom0}).  All three \ion{H}{1}
concentrations are located at the ends of the stellar disks with the C concentration in the ``interaction region''
between the stellar disks.  These three \ion{H}{1} concentrations are also the most remarkable features near
the main body of the galaxies in the channel maps (Fig.~\ref{h1chmap}), especially the C concentration.  In
the higher resolution map (Fig.~\ref{h1unifmom0}), the C concentration is marginally resolved by the
$12\arcsec$ beam and appears to extend toward the north-eastern direction.  The compact W concentration
remains unresolved in the high resolution map.

The \ion{H}{1} gas mass is estimated using the formula \citep{roberts75}:
\begin{displaymath}
M_{\rm H I} (M_{\sun}) = 2.36 \times 10^5 D^2 \int Sdv,
\end{displaymath}
where D is the distance in Mpc, and $\int Sdv$ is the \ion{H}{1} integrated flux in Jy km $\rm s^{-1}$.   
The \ion{H}{1} gas masses of the whole NGC 6670 system, the W tail, and the NE tail obtained from the 
low resolution channel maps are
$1.6\times10^{10}M_{\sun}$, $3.3\times10^9 M_{\sun}$, and $8.1\times10^8 M_{\sun}$, respectively.
The \ion{H}{1} mass within the C and W concentrations obtained from the high resolution channel maps is
$2.0\times10^9 M_{\sun}$ and $1.4\times10^9 M_{\sun}$, respectively.  We further estimate the \ion {H}{1}
peak and averaged column density in the C and the W concentration in the high resolution integrated intensity
map (made with 1.5 $\sigma$ flux clipping).  The results are listed in Table~\ref{tab1}.  Because the W
concentration is unresolved and the C concentration is only slightly larger than the beam, the measured peak
column densities should be regarded as lower limits.

As mentioned in \S~\ref{optresults}, the C concentration is located in the place where the heaviest dust 
obscuration occurs.  The visual extinction caused by the C concentration can be estimated with the relation 
$A_V \sim n_{\rm HI}/2 \times 10^{21}\rm cm^{-2}$.  Using the peak \ion{H}{1} column density listed in
Table~\ref{tab1}, we estimate the visual extinction to be $\sim 1$ magnitude.  This is a lower limit because
individual \ion{H}{1} clouds in the C concentration are unresolved and may have column densities higher 
than the value used here.  Furthermore, molecular gas could also be present.  It seems likely that the C
concentration contributes to, if not dominates, the extinction in the eastern portion of the disk of 
NGC 6670W.  If this is true, the fact that there is not obvious extinction in the western half of the NGC 
6670E disk would imply that NGC 6670W is in the background, relative to NGC 6670E.  We will discuss 
the geometry of this interacting system in details in \S~\ref{interaction2}.

\subsubsection{\ion{H}{1} Kinematics in NGC 6670}\label{h1motion}

In Figure~\ref{pvplots}, we present \ion{H}{1} position-velocity ($p$-$v$) diagrams along the major axes
of stellar/molecular disks and along a slice joining the two nuclei.  For comparison, we also overlay the CO
$p$-$v$ diagrams (see \S~\ref{comotion}) along the same slices used for the \ion{H}{1}.  On the large scale, 
the motion of \ion{H}{1} near the two stellar disks appears as a single rotating system.  Unlike in the CO 
$p$-$v$ diagrams, it is hard to see two distinct \ion{H}{1} systems (Fig.~\ref{pvplots}c).  This is also true 
in the channel maps (Fig.~\ref{h1chmap}).  In the channel maps, from $v=8345$ to 8760 km $\rm s^{-1}$, 
the location of \ion{H}{1} emission changes systematically from the west to the east.  An explanation to 
this is that the \ion{H}{1} gas originally in the two disks is now merging into a single dynamical system.   
On the other hand, the CO velocity field suggests the galaxies have similar spin directions
(\S~\ref{comotion}).  It is also possible that the apparent global \ion{H}{1} kinematics is just a result caused 
by chance matching of the velocities of the galaxies.  

The \ion{H}{1} near the disk of NGC 6670W (Figure~\ref{pvplots}b) shows a constant velocity gradient
over the length of the optical disk (between $-20\arcsec$ and $+20\arcsec$).  The \ion{H}{1} emission 
along the velocity gradient is weakest near the systemic velocity of NGC 6670W, 8570 km $\rm s^{-1}$, 
which is determined by the CO velocity field (\S~\ref{comotion}).  The stronger \ion{H}{1}
emission is located near the ends of the optical disk (the C and W concentrations).  
The constant velocity gradient and the edge-brightened morphology
in the $p$-$v$ diagram suggest that the \ion{H}{1} gas is in a rotating ring-like structure around
NGC 6670W.  The rotation velocity of the \ion{H}{1} ring is $\sim$150 km $\rm s^{-1}$, slightly lower
than the rotation velocity of the molecular disk which is $\sim175$ km $\rm s^{-1}$ (\S~\ref{comotion}).
The edge-on \ion{H}{1} ring partially explains the existence of the C concentration and the W concentration.

The \ion{H}{1} $p$-$v$ diagram for NGC 6670E (Fig.~\ref{pvplots}a) has a more complex appearance
than that for NGC 6670W.  There is a lack of \ion{H}{1} emission near the center of NGC 6670E at the
systemic velocity suggested by the CO velocity field, around $v=8730$ km $\rm s^{-1}$
(\S~\ref{comotion}).  Most of the  \ion{H}{1} emission at the center position comes from a small peak 
centered between $-2\arcsec$ and $+8\arcsec$ at 8630 km $\rm s^{-1}$.  This gas cannot be in a circular 
orbit and its velocity suggests either 100 km $\rm s^{-1}$ infall or outflow relative to the center of NGC 6670E.
Also present in the \ion{H}{1} $p$-$v$ diagram of NGC 6670E are the C and E concentrations.  The E
concentration appears to roughly match the CO velocity in the eastern part of NGC 6670E.  Two velocity 
components of the C concentration at $v=8650$ and 8715 km $\rm s^{-1}$ appear to associate with the 
western disk of NGC 6670E and also the eastern disk of NGC 6670W (Fig.~\ref{pvplots}b,c).  Overall, 
the \ion{H}{1} velocity distribution in 
NGC 6607E is neither like a rotating ring (e.g., \ion{H}{1} in NGC 6670W), nor like a rotating disk 
(e.g., CO in NGC 6670E).  This suggests that the \ion{H}{1} disk of NGC 6670E has almost been destroyed.
The clear deviation from circular rotation of \ion{H}{1} in NGC 6670E is another strong indication 
of a tidal interaction between the galaxies besides the apparent \ion{H}{1} tidal tails.

The C concentration shows two velocity components ($v=8715$ and 8650 km $\rm s^{-1}$) that can be
interpreted as belonging to either or both galaxies.   It is interesting to note that both the E and 
W concentrations have velocities close to the CO velocities in the same regions.  On the other hand, 
only the weaker 8650 km $\rm s^{-1}$ component of the C concentration appears to match the CO 
velocity of the western part of NGC 6670E.  The strongest component (8715 km $\rm s^{-1}$) has 
a velocity between the CO velocities of the eastern part of NGC 6670W (8740 km $\rm s^{-1}$) and the 
western part of NGC 6670E (8660 km $\rm s^{-1}$).  This makes the outer \ion{H}{1} in both galaxies 
appear to have lower circular velocities than the molecular gas in the inner disks.   

As mentioned earlier in this section, the global apparent \ion{H}{1} kinematics could be explained by the 
merge of \ion{H}{1} disks or by matching up the \ion{H}{1} velocities of the disks by chance.  The apparent
lower \ion{H}{1} rotation velocities in both disks and the existence of the C concentration could also be 
explained by these two scenarios.  First, if the \ion{H}{1} disks were interacting with each other and merging 
into a single system, the C concentration might result from a direct interaction between the \ion{H}{1} disks.  
In other words, cloud-cloud collisions, ram pressure, or tidal stripping in the prograde-prograde 
 (\S~\ref{comotion}) interaction may be forming a large gas reservoir in the region between the galaxies.  The
collisions may slow down the velocities of the \ion{H}{1} clouds between the disks and produce the
observed \ion{H}{1} velocity profiles.   In this scenario, the C concentration is a true concentration of 
\ion{H}{1} clouds in space.   On the other hand, it is also possible that the two \ion{H}{1} disks are spatially
separated along the line of sight.   Instead of direct collisions, the lower \ion{H}{1} rotation velocities might 
result from loss of angular momentum due to a tidal torque  from the interaction.  
Therefore, in the region between the galaxies, the \ion{H}{1} clouds from the two disks have velocity 
difference smaller than that suggested by CO rotation curves.  In this scenario, the C concentration is a
superposition of two independent structures along the line of sight.

The most important difference between these two scenarios is whether or not the two \ion{H}{1}
components of the C concentration are approaching each other.  As mentioned in \S~\ref{h1}, it is suggested
that NGC 6670W is behind NGC 6670E.  Consequently, the CO velocities of the eastern part of NGC 6670W and 
the western part of NGC 6670E suggest that the \ion{H}{1} clouds from the two disks in the C concentration 
must be moving away from each other if there is no strong interaction between the \ion{H}{1} clouds.  On 
the other hand, a direct interaction requires the \ion{H}{1} clouds to be approaching each other.  These two 
possibilities can be distinguished if we can tell which \ion{H}{1} component belongs to which galaxy.  
However, this is not possible no matter in the $p$-$v$ diagrams or the channel maps in our current 
\ion{H}{1} data of very limited resolution.    Higher resolution \ion{H}{1} observations may help to answer this.

\subsection{CO(1--0) Results}\label{coresult}
\subsubsection{Distribution of the Molecular Gas}\label{codist}
Our BIMA CO(1--0) observations reveal the distribution of the molecular gas in NGC 6670.  In 
Figure~\ref{cochmap}, we present the CO(1--0) channel maps of NGC 6670 at $2\farcs7\times2\farcs1$ 
resolution.  In Figure~\ref{comoms}, we present the CO integrated intensity at resolution of 
$5\farcs4\times4\farcs6$ (Fig.~\ref{comoms}a, hereafter the low resolution) and 
$2\farcs7\times2\farcs1$ (Fig.~\ref{comoms}b, hereafter the high resolution).  The CO emission
in NGC 6670 is distributed along the inner part of the optical disk in each galaxy.  No CO emission was
detected in the \ion{H}{1} tail regions outside the optical disks.  Little CO emission is detected from the
region where the stellar disks overlap (Fig.~\ref{cochmap}, $v=8660$ and 8680 km $\rm s^{-1}$).  Near the 
center of NGC 6670W, the CO emission appears to extend slightly in the direction perpendicular to the disk.  
In NGC 6670E, the CO distribution is more distorted and deviates from the symmetric disk morphology.  Both 
the CO morphology and the \ion{H}{1} kinematics shown in Figure~\ref{pvplots} suggest a stronger tidal 
perturbation on the disk of NGC 6670E.

In Figure~\ref{comoms}b,  the distribution of dust absorption as suggested by the $B$-band image seems 
to correlate with the molecular gas distribution.  Therefore, it seems plausible that the molecular gas in the 
eastern part of the NGC 6670W disk contributes to the extinction in this region.  On the other hand, it is 
suggested in Figure~\ref{h1unifmom0} that the \ion{H}{1} C concentration may also contribute to the 
extinction.  The $\rm H_2$ column density in this region is an order of magnitude higher than the lower limit 
of the peak \ion{H}{1} column density.  However, comparing to the \ion{H}{1}, the molecular clouds in 
NGC 6670W are distributed more symmetrically over the whole disk.  It is unlikely that all the $\rm H_2$ 
in the eastern part of the NGC 6670W disk contributes to the extinction because some molecular clouds in 
the edge-on disk may be behind the stars.  It is not clear in our observations which gas 
component in the eastern part of the disk of NGC 6670W has more contribution to the extinction.

The CO(1--0) images show that the large molecular reservoirs are located in the central regions of the disks.
We measure the CO integrated flux in the inner regions of the two galaxies from the high resolution
channel maps.  We find that 56\% of the CO flux from NGC 6670E is concentrated in the inner 4 kpc region.
In NGC 6670W, 48\% of its CO flux comes from the inner 4 kpc region.  The molecular gas distribution in
both NGC 6670W and NGC 6670E is clearly concentrated toward the central region, especially knowing that 
the extent of the stellar disks of both galaxies is $\gtrsim 20$ kpc.  

The total CO integrated flux is determined using the high resolution channel maps (Fig.~\ref{cochmap}).
The total flux is 65.4 Jy km $\rm s^{-1}$ for NGC 6670E and is 149.7 Jy km $\rm s^{-1}$ for NGC
6670W.  The total flux of the NGC 6670 system measured by the BIMA array is the same, within the errors, 
as the single dish value of 220 Jy km $\rm s^{-1}$ \citep{gao96,gao99}.  The shapes of the integrated 
CO spectra of NGC 6670E and NGC 6670W (Fig.~\ref{cospect}) are also similar to the single dish spectra of 
\citet{gao99}.  These comparisons suggest that the interferometric observations did not systematically miss 
CO emission.

We estimate the molecular gas mass from the integrated CO(1--0) flux using the empirical
relation \citep{scoville87}:
\begin{displaymath}
M_{\rm H_2}(M_{\sun}) = 1.18 \times 10^4 D^2 \int Sdv,
\end{displaymath}
where $D$ is distance in Mpc and $\int Sdv$ is CO integrated flux in Jy km $\rm s^{-1}$.
Here the Galactic CO-to-$\rm H_2$ conversion factor
$X \equiv N(\rm H_2)$/$I_{\rm CO}=3.0 \times 10^{20}$ $\rm cm^{-2}$ \rm (K km $\rm s^{-1})^{-1}$
(see \citet{young91} for a review) is assumed for convenient comparison with previous results.   We note that
this conversion factor may lead to an overestimate of the molecular gas mass if the galaxy is undergoing
starbursts (see e.g., \citet{maloney88,bryant99}).  The estimated total molecular mass is $1.1 \times 10^{10}$
$M_{\sun}$ for NGC 6670E and is $2.5 \times 10^{10}$ $M_{\sun}$ for NGC 6670W.

We estimate the peak edge-on molecular column density in the inner 2 kpc region of each galaxy
using the high resolution integrated intensity map.  The results are listed in Table~\ref{tab2}.  It is difficult 
to obtain a precise face-on column density for these two galaxies.  This is mainly because of the irregular 
CO morphology in the outer disks and therefore there is a large uncertainty in determining the
outer disk contribution to the edge-on column density.  To simply the estimation, we assume that the CO flux 
measured in the central 2 kpc regions in the image has no outer disk contribution and all comes from the 
inner disks.  The estimated central face-on column density is close to the edge-on values ($\sim 10^3$
$M_{\sun}$ $\rm pc^{-2}$).  The molecular hydrogen column density is large compared with that of the 
Milky Way's disk (4--20 $M_{\sun}$ $\rm pc^{-2}$) and center (40--300 $M_{\sun}$ $\rm pc^{-2}$
\citep{sanders84}) but small compared with that of the LIRGs at the more advanced merging stages
(typically greater than $10^4$ $M_{\sun}$ $\rm pc^{-2}$, see e.g., \citet{bryant99,yun95,solomon92}).

Combining with the results of our VLA \ion{H}{1} 21 cm observation, the total gas content (atomic +
molecular) in this system is $\sim 5.3\times10^{10}M_{\sun}$ and the global ratio of $M_{\rm H_2}$ to
$M_{\rm HI}$ is 2.2.  According to the study of \citet{young89}, Sab and Sb spiral galaxies have a ratio
of $M_{\rm H_2}/M_{\rm HI}$ around 2 and spiral galaxies later than Sc have a ratio smaller than 1.  The
optical morphology of NGC 6670 suggests that the two disks are late type spiral galaxies.   The slightly 
high global ratio of $M_{\rm H_2}$ to $M_{\rm HI}$ for NGC 6670 might be interpreted as due to an 
enhanced conversion of atomic to molecular gas.  Furthermore, roughly 25\% of \ion{H}{1} gas in NGC 
6670 is in widely extended tidal tails and roughly 22\% is additionally concentrated outside the disks. 
The ISM in the inner galactic disks is predominately molecular.

\subsubsection{Kinematics of the Molecular Gas}\label{comotion}
The molecular gas in the disks of NGC 6670 forms two independent dynamical
systems.  In the channel maps (Fig.~\ref{cochmap}), the location of CO emission in each disk moves
systematically from the west to the east.  This is also clear in the radial velocity map (Fig.~\ref{comom1})
that the velocity in each disk increases from the western end to the eastern end.  In particular, the velocity
of the stellar fuzzy feature in the southeastern end of NGC 6670E (\S~\ref{optresults})
is consistent with the velocity distribution of the main body of NGC 6670E, which is most clear in 
Figure~\ref{pvplots}a ($12\arcsec$ to the east of the center) and in Figure~\ref{cochmap} ($v=8760$ to 8820
km $\rm s^{-1}$).  Together with the CO morphology shown in Figure~\ref{comoms}a, the consistency of 
velocity suggests this fuzzy feature is part of the disk of NGC 6670E and not another galaxy.

The continuous change of CO velocity in the two galaxies as mentioned above and the double-peak line 
profiles of NGC 6670W (Fig.~\ref{cospect}) suggest that the two molecular disks are both rotating.  
The CO $p$-$v$ diagram in Figure~\ref{pvplots}b shows the inner rotation curve of NGC 6670W.  
The CO emission has a relatively uniform distribution along a constant velocity gradient of $\sim60$ km 
$\rm s^{-1}$ $\rm arcsec^{-1}$ between 8320 and 8860 km $\rm s^{-1}$ in the central $6\arcsec$ (3.6 kpc) 
of the disk.  In addition, in the channel maps (Fig.~\ref{cochmap}), one can find CO emission from the 
center of NGC 6670W in almost every channel between $v=8300$ and 8840 km $\rm s^{-1}$.  The 
relatively uniform CO brightness over 540 km $\rm s^{-1}$ suggests that the concentrated CO emission at the 
center of NGC 6670W in the integrated intensity map is the sum of CO in velocity space, rather than a
concentration in true space. 

In the outer regions of NGC 6670W, the CO rotation curve appears to become flat.  The velocities of the flat 
portion of the rotation in the radial velocity map (Fig.~\ref{comom1}) are 8400 and 8740 km $\rm s^{-1}$ on 
the western and eastern side of the NGC 6670W disk respectively.  This provides that the systemic velocity of
NGC 6670W is 8570 km $\rm s^{-1}$ and the maximum rotation velocity is 170 km $\rm s^{-1}$.

The constant velocity gradient in the inner 4 kpc of NGC 6670W has a velocity range which exceeds those
observed in the outer portions of the galaxy.  This suggests that some of the molecular gas in the inner part of
NGC 6670W are on non-circular orbits.  We suggest that the most natural explanation for this is a central
molecular bar with a projected spatial extent of $\lesssim 3.5$ kpc at a small angle ($<45\arcdeg$) 
with respect to the line of sight.  Similar morphology in $p$-$v$ diagrams can be found in several other barred
spiral galaxies, such as NGC 5746, NGC 5965 \citep{kuijken95}, and NGC 4102 \citep{jogee96}.  Theoretical
studies have shown that stellar and gaseous bars may play an important role in driving gas into the central 
regions of galaxies \citep{schwarz84,athana92,barnes91}.  \citet{mihos96} further showed that the angular
momentum of gas can be removed rapidly along a bar permitting intense gas inflow to take place.  As suggested
by the integrated CO flux, the central molecular bar in NGC 6670W may contain more than 
$\sim 1.2\times10^{10} M_{\sun}$ of molecular gas.  If the rapid inflow of molecular gas predicted by the
simulations takes place in NGC 6670 as the merging process progresses, the center of NGC 6670W will become 
a powerful starburst region with a IR luminosity far exceeding the present value .

The CO $p$-$v$ diagram of NGC 6670E (Fig.~\ref{pvplots}a) suggests the motion of the molecular gas is
disturbed, as in the case of the \ion{H}{1} gas (\S~\ref{h1motion}).
As shown in Figure~\ref{pvplots}a, the $p$-$v$ diagram of CO emission along NGC 6670E's major axis
appears less symmetric than that of NGC 6670W.  While the rotation curve appears to become flat on the
eastern end of the disk, this is not as clearly observed on the western side.  Thus the determination of the
systemic and maximum rotation velocity is less certain.  We use the averaged velocity in the innermost region
to estimate the systemic velocity of NGC 6670E, which is 8730 km $\rm s^{-1}$.  Our best estimate of the
maximum rotation speed is based on the velocity difference along the velocity gradient in the inner portion at
the $4\sigma$ level.  The resultant value is $\sim 100$ km $\rm s^{-1}$.  We note that, unlike NGC 6670W,
the CO emission near the center of NGC 6670E is concentrated in both velocity and spatial extent
(Fig.~\ref{pvplots}a).  Thus the molecular concentration in the center of NGC 6670E seems more like a 
true concentration.


Figure~\ref{comom1} shows that the spin vectors of the two galaxies are both roughly toward the south.
It also shows that, in this system, NGC 6670E is redshifted relative to NGC 6670W by 160 km $\rm s^{-1}$.  
If the transverse velocities of both galaxies are small compared with the radial velocities, then the orbital vector
of this system would also be toward the south.  Such a configuration indicates a prograde-prograde
interaction.  Prograde-prograde encounters have been shown to be the most efficient at producing long tails
\citep{toomre72} and to have stronger gas inflow toward the nuclei of mergers \citep{mihos96}.
Although we cannot determine the true orbital motion of each galaxy without knowing the transverse velocities,
a prograde-prograde geometry will give the best consistent explanation to the observed \ion{H}{1}
distribution and kinematics.  We will further discuss this in \S~\ref{interaction2}.

\subsection{20 cm Radio Continuum Results}\label{20cmresult}
The VLA 20 cm radio continuum map at $3.1\arcsec$ resolution is shown in Figure~\ref{20cm}.  The 
one-dimensional profiles of 20 cm flux density and CO integrated intensity are also plotted in
Figure~\ref{profiles} for comparison.  Globally, the radio continuum emission is well correlated with 
both the CO emission and the optical emission.  Two stronger radio continuum peaks appear at the eastern end 
of NGC 6670E and the western end of NGC 6670W (Fig.~\ref{profiles}).  The peak at the western end of 
NGC 6670W appears correlated with a bright region in the $B$-band image which may be a recent star 
forming region.  The peak at the eastern end of NGC 6670E  is roughly coincident with the heavily obscured
region and may indicate recent embedded star formation activity.

The 20 cm emission in the center of each galaxy is marginally resolved by the $3\farcs1$ circular beam
along the major axis of each galaxy (Fig.~\ref{profiles}).  This 
suggests that the emission in each nucleus does not originate from a central point source such as an AGN, 
but originates from extended star forming regions.  A similar suggestion for the energy source of the nuclear
regions of NGC 6670 had been made based on the ISO mid-infrared observations \citep{hwang99}.  The 
peak brightness in each nucleus is 10.3 mJy $\rm beam^{-1}$ (607 K) in NGC\,6670E and 9.7 mJy
$\rm beam^{-1}$ (571 K) in NGC\,6670W.  The bright emission in the central regions of the galaxies implies
that the strongest star formation activity occurs in the nuclear regions.  Numerical simulations of interacting
galaxies predict gas inflows into the nuclear regions and therefore anticipate a similar concentrated active star
formation activity in the nuclear regions \citep{barnes91,mihos93,mihos96}.

Roughly 65\% of the 20 cm flux of NGC 6670E and 50\% of the flux of NGC 6670W originate
from the inner 4 kpc region of each disk. The total flux density is 22.3 mJy for NGC 6670E, 35.6 mJy for NGC
6670W, and 60.4 mJy for the whole system.   Furthermore, the 20 cm flux density and the far-infrared (FIR)
flux density of NGC 6670 appear to obey the FIR--radio correlation of normal galaxies (galaxies without AGNs).
The logarithmic FIR to radio flux density ratio defined by \citet{condon91a} is
\begin{displaymath}
q\equiv log [(FIR/3.75 \times 10^{12}\rm Hz)/\it S_{\rm 1.49GHz}],
\end{displaymath}
where the 20 cm flux density $S_{1.49\rm GHz}$ is in W $\rm m^{-2}$ $\rm Hz^{-1}$.  The quantity
$FIR$ is defined as
\begin{displaymath}
FIR(\rm W~m^{-2})\equiv 1.26\times10^{-14}(2.58\it S_{\rm 60\mu m}\it +S_{\rm 100\mu m}),
\end{displaymath}
with the $60\mu m$ flux density $S_{60\mu m}$ and the $100\mu m$ flux density $S_{100\mu m}$ in Jy.  
Radio and FIR selected starbursts as well as optically selected spiral and irregular galaxies have a very narrow
$q$-distribution ($\sigma_q=0.19$) centered on $<q>=2.34$.  On the other hand, galaxies with radio emission
powered by AGNs generally have $q<2$ \citep{condon92}.  Given $S_{60\mu m}=8.25$ Jy,
$S_{100\mu m}=15.2$ Jy \citep{moshir90}, and the measured $S_{1.49\rm GHz}$, the value we find for $q$
in NGC 6670 is 2.3.  This is a strong indication that all the radio emission and infrared luminosity from
NGC 6670 arise from star formation rather than AGN activity.

\section{Discussion}

\subsection{The Dynamics of the Interaction}\label{interaction}
Our observations of NGC 6670 present the first example of an interacting system in which the \ion{H}{1} 
disks have been highly perturbed but the individual stellar and molecular disks show relatively little distortion.  
The tidal tails of NGC 6670 only contains $\sim25\%$ of the total \ion{H}{1} gas.  
NGC 6670 is among the few cases of interacting systems found to have a long (90 kpc) but less massive
\ion{H}{1} tail without obvious stellar counterparts.  An extremely deep optical imaging is required to 
confirm the absence of stellar tails.  Another system with similar \ion{H}{1} and optical properties is 
NGC 3226/3227 \citep{mundell95}.  The \ion{H}{1} morphology and kinematics in NGC 6670 are also 
strikingly different from those of the molecular or stellar components.  

In this section, we aim to explain the distribution and motion of atomic and molecular gas and to understand 
the dynamics of the interaction in this system.   While there is no specific simulation of NGC 6670, 
comparisons of our observation with the study of gravitational interaction of galaxies by 
\citet{toomre72}(hereafter TT72) can allow a fair amount of understanding of the interaction history 
and the orbital geometry of NGC 6670.

\subsubsection{Masses of the Galaxies}\label{mdyn}
The dynamical mass within radius $r$ in a galaxy can be estimated using
\begin{displaymath}
M(r)=r(v/\sin{i})^2/\rm G,
\end{displaymath}
where $i$ is the inclination angle of the disk and $v$ is the projected rotation velocity of the disk.  We chose
the outermost radius of the stellar disk measured in the $K^{\prime}$-band image for $r$, which is 12 kpc
for NGC 6670W and 10 kpc for NGC 6670E.  The rotational velocities are determined from the CO radial
velocity (\S~\ref{comotion}).  Because $i\sim90\arcdeg$ for both galaxies, the dynamical masses are close to 
$8.1 \times 10^{10} M_{\sun}$ for NGC 6670W and $2.3\times10^{10} M_{\sun}$ for NGC 6670E.  

In $K^{\prime}$-band, the flux of NGC 6670W is 1.54 times greater than that of NGC 6670E (D.-C. Kim
2000, private communication).  If the two galaxies have identical $K^{\prime}$-band mass-to-light ratio and 
the dynamical mass of NGC 6670W should be reliable as its rotation is less perturbed, the $K^{\prime}$-band 
light ratio implies the mass of NGC 6670E to be $5.3\times10^{10} M_{\sun}$, about twice the estimate above.
The large difference between NGC 6670E's dynamical mass and luminosity mass may suggest that the rotation 
of CO in NGC 6670E is too highly disturbed to be a good indicator of the dynamical mass.  

\subsubsection{Comparison of the Molecular and Atomic Gas Motion}
In interacting galaxy systems, the response of atomic and molecular gas to the tidal perturbation may be 
different.  As the \ion{H}{1} gas is generally distributed in the outer regions of spiral galaxies further beyond 
the stellar disks \citep{broeils94}, it is less gravitationally bound to its host galaxies and is closer to the 
perturbing galaxy.   Therefore the \ion{H}{1} gas is more sensitive to tidal perturbation than the molecular gas 
that is mostly found in the inner regions of galaxies.  \ion{H}{1} clouds can more easily escape to form 
tails, bridges, or shells during the interaction.  

The study of gas clouds in optically selected interacting
galaxies (i.e., galaxies with stellar tails, bridges or other morphological peculiarity) have concentrated on the
\ion{H}{1} gas (e.g., the study of \ion{H}{1} in a ``Toomre Sequence'' by \citet{hibbard96}).  On the
other hand, the study of molecular gas properties in merging galaxies have concentrated on advanced or
late-stage mergers (e.g., \citet{sanders88,scoville91,downes98}).  To date, there have been few studies
containing observations of both molecular and atomic gas in early-stage interacting systems.

In NGC 6670, the response of molecular and atomic gas to the interaction is completely different even in the
inner disk regions.  While the molecular disks are still well separated and keep their identity of disk rotation,
we can not tell two distinguished rotating \ion{H}{1} systems from the $p$-$v$ diagram 
(Figure~\ref{pvplots}c).   The \ion{H}{1} disks are probably merging into a single rotating system 
dominated by the more massive galaxy NGC 6670W.  This same trend is also observed in other 
interacting systems such as NGC 520 \citep{hibbard96} and Arp 299 \citep{hibbard99}, where the 
\ion{H}{1} disks had already merged into a single system but the merging of stellar disks is still ongoing.

In NGC 6670W, the \ion{H}{1} gas is likely in a rotating ring-like structure that has a rotation velocity slightly
lower than that of the molecular disk (Fig.~\ref{pvplots}b).  Much of the \ion{H}{1} emission comes from the
two outer ends of NGC 6670W's stellar disk (the C and W concentrations).  On the other hand, the strongest
molecular emission in NGC 6670W is located in the circum-nuclear regions.  The concentrated molecular 
clouds in its center appear to be in a bar elongated roughly along the line of sight, and the outer molecular 
disk appears to have a flat rotation curve, as shown in Figure~\ref{pvplots}b.  Overall, the motion and 
distribution of the molecular clouds seem less perturbed in NGC 6670W, compared with those in NGC 6670E.

The CO(1--0) morphology, the CO velocity field, and the \ion{H}{1} velocity field all suggest that the
gravitational perturbation acts stronger on NGC 6670E than NGC 6670W.   This is probably because
NGC 6670W is more massive than NGC 6670E.  The rotation of  \ion{H}{1} in NGC 6670E is
almost unidentifiable in the \ion{H}{1} $p$-$v$ diagram along its major axis (Fig.~\ref{pvplots}a).
The \ion{H}{1} $p$-$v$ diagram for NGC 6670E also shows some non-circular motion
of \ion{H}{1}, as described in \S~\ref{h1motion}.  While the original \ion{H}{1} disk around
NGC 6670E has nearly been destroyed, the molecular disk of NGC 6670E has not. As suggested by the 
CO $p$-$v$ diagram of NGC 6670E, the molecular clouds in the innermost of the galactic disk still rotate 
around the nucleus, while the flat outer part of its rotation curve on the western side appears disrupted.  
We also find that the CO rotation of NGC 6670E is highly perturbed by comparing its luminosity mass and 
dynamical mass.  In addition, the CO emission from NGC 6670E seems more
concentrated toward the nuclear region than in NGC 6670W.  All the above may result from the tidal 
perturbation produced by the interaction.

\subsubsection{Orbital Geometry and Interaction History}\label{interaction2}
There are several important results in the study of TT72 that can be used to infer the orbital geometry and the 
interaction history of NGC 6670.  1) Only a prograde passage with orbital inclination $i\lesssim30\arcdeg$
with respect to the spin plane of the perturbed galaxy can produce long tidal tails.  2) Tidal tails can only be
generated after the galaxies have passed their pericenters.  3) During the prograde encounter, only particles 
orbiting their host galaxies with radii greater than 0.5 pericentric distance can be ejected to form tidal tails if the 
interacting pair have comparable masses.

Our VLA observation reveals an \ion{H}{1} tail of 90 kpc in extent.  According to the first point from TT72,
the interaction in NGC 6670 must be prograde for at least one of the galaxies.  The CO velocity field reveals
that the galaxies have similar direction of spin.  Therefore the interaction should be a prograde-prograde
encounter.  This is also suggested by the CO systemic velocity of each galaxy and the global velocity of the
\ion{H}{1} disks (Fig.~\ref{pvplots}c).  Furthermore, besides the short NE tail extending in the
N-S direction, most of the \ion{H}{1} in this system is distributed along the E-W direction. This suggests that
the orbital plane of the interacting galaxies in NGC 6670 lies nearly parallel to the plane defined by the line
of sight and the E-W direction, and the transverse velocities in the N-S direction are small.  These are all 
consistent with the hypothesis of a prograde-prograde interaction.

The second point from TT72 mentioned above states that the galaxies have already finished their first approach
and have been interacting with each other for at least long enough to produce a 90 kpc tail.  We can estimate 
the pericentric distance of this system by noting that there are no significant stellar tails detected in optical and 
NIR images.   According to the third point mentioned from TT72, if there were significant stellar components
orbiting either galaxy with radii greater than 0.5 pericentric distance, we should see a stellar tail in this system.  
This gives a lower limit for the pericentric distance.   Therefore the extent of NGC 6670W's stellar disk in the
optical images is the lower limit for the pericentric distance, $\sim 25$ kpc.  With such a large impact 
parameter, only the extended \ion{H}{1} disk has formed tails.  This also explains why NGC 6670 has 
relatively less \ion{H}{1} distributed along the tails compared with other systems like NGC 4676 or Arp 
244 (\S~\ref{h1}) which may have had smaller pericentric distances in their most recent encounters.  We 
conclude that the difference in appearance between the \ion{H}{1} and stellar-plus-molecular components 
in NGC 6670 can be ascribed to a relatively large impact parameter and the different initial distribution of 
those components.  

It is worth to point out that the present projected separation of the galaxies (16 kpc) is less than the lower 
limit of the previous pericentric distance (25 kpc).  This suggests either that the separation of the galaxies 
along the line of sight is larger than $\sim 20$ kpc, or that the galaxies are during their next close encounter 
which should have a smaller pericentric distance.   The following discussion prefers the second case.

NGC 6670 (and also NGC 3226/3227, \citep{mundell95}) demonstrates that \ion{H}{1} tails can form
independently of stellar tails.  The formation of \ion{H}{1} tails without stellar counterparts might be a 
common phenomenon for early stage mergers with large pericentric distances.  On the other hand, in this kind 
of interacting systems, the stellar tails can still form in their later stage of merging.  Since the \ion{H}{1}
tails and probably also the disrupted dark halo are carrying part of the galaxies' angular momentum, the 
galaxies should have a smaller pericentric distance during their next close encounter than the previous one.  
If there are stars located outside the radius of 0.5 pericentric distance during the next encounter, they will
likely be ejected to form stellar tails.  It is also possible to form new minor \ion{H}{1} tails during future
encounters.  Such minor tidal features forming in later encounters can be found in the simulations.  Due to 
the different initial distribution of stellar and gaseous components, different tidal features made of stars or 
gas could be generated after a few orbital periods.  By the same token, the observed offset or morphological
inconsistency between stellar and gaseous tails/bridges in many other interacting systems (e.g., the M81 
group, \citet{yun94}; Arp 299, \citet{hibbard99}) might be explained by the different initial distributions of 
the different components.  Simulations by \citet{mihos01} had already verified this suggestion.

The interaction age of NGC 6670 can also be estimated if we know the true spatial velocity ($V$) of the tails.
Given the projected 90 kpc extension of the W tail, the time elapsed since this tail began forming must be
at least 90 kpc/$V$.  Consider that the tails are escaping particles that became nearly gravitationally unbound
to the host galaxies due to the tidal force of the interaction (TT72).  Thus we assume that the escaping
particles were not accelerated to higher velocities during the encounter but have velocities comparable to
their original velocities prior to the interaction.  Therefore, the value of $V$ can be estimated as the rotation
velocity plus orbital velocity of a galaxy.  We use the value $\triangle V/2 + v$ for $V$, where $v$ is
the rotational velocity of W tail's parent galaxy and $\triangle V$ is relative radial velocity of the galaxies.
Given that the rotational velocities of the galaxies are 100 and 170 km $\rm s^{-1}$ and $\triangle V=160$
km $\rm s^{-1}$, the possible $V$ ranges between 180 and 250 km $\rm s^{-1}$.  Thus the age of the tail is 
$\sim 4\times10^8$ yr.  The age of the tail can perhaps be used to indicate the age of interaction.  
Similar dating methods were used on other interacting systems (e.g., \citep{hibbard99,murphy2000}).
The interaction age of the advanced merger Arp 299 was estimated to be $7.5 \times 10^8$ yr 
\citep{hibbard99}.  This is almost twice the interaction age of NGC 6670.  Although there is an uncertainty 
(probably a factor of 2) with this kind of age estimate, it is still plausible that NGC 6670 is in an early
stage of merging.

In addition to the age of interaction, we can also estimate the orbital time scale of this system using
$\pi R/\triangle V$ where $R$ is projected nuclear separation and $\triangle V$ is relative radial velocity
between the galaxies.   Given $R=16$ kpc, the orbital time scale is $\sim 3\times10^8$ yr.  This time scale
is comparable with the interaction age.   It is therefore possible that the second approach of the galaxies are 
now ongoing if the galaxies are in gravitationally bound orbits.  Furthermore, as described in \S~\ref{h1}, the
\ion{H}{1} C concentration may cause the extinction in the eastern side of the disk of NGC 6670W.  
This implies that the C concentration is in front of NGC 6670W along the line of sight.  Moreover, 
the lack of obscuration in the western part of NGC 6670E suggests that NGC 6670E is in front of the C
concentration and therefore is in front of NGC 6670W.  If this is the case, the galaxies must be approaching 
to each other because NGC 6670W is blueshifted relative to NGC 6670E.

We can estimate the ratio of the kinetic and potential energies of this system to test whether the two
disks are gravitationally bound.  The ratio is
\begin{displaymath}
T/|U| \approx (M_r \triangle V^2 /2)/(G M_E M_W /R)=R\triangle V^2/(2G(M_E+M_W)),
\end{displaymath}
where $M_r$ is the reduced mass of the system, $\triangle V$ is the relative velocity and $R$ is the nuclear
separation of the galaxies with masses $M_E$ and $M_W$.  For NGC 6670W, we use the dynamical mass,
$8.1\times10^{10}M_{\sun}$.  For NGC 6670E, we use the luminosity mass $5.3\times10^{10} M_{\sun}$
as described in \S~\ref{mdyn}.   Using the projected nuclear separation (16 kpc) for $R$ and the radial
velocity difference (160 km $\rm s^{-1}$) for $\triangle V$, we estimate the ratio $T/|U|$ to be 0.35.  Because
both $\triangle V$ and $R$ used here are lower limits, this estimate $T/|U|$ is also a lower limit.  Although we
cannot obtain an upper limit for $T/|U|$, it is reasonable to conclude that the galaxies are gravitationally bound
($T/|U|<1$) for three reasons.  1) The masses we use are lower limits because they only  include material
within the stellar distribution.  The inclusion of material outside the extent of the stellar disks would
make the ratio $T/|U|$ smaller.  2) As described in the previous paragraph, the interaction age is comparable 
with the orbital time scale.  If the galaxies were unbound, they would have a much larger separation than the
present projected separation after an orbital time scale since the first encounter.   The relatively small projected
separation suggests the galaxies to be bound, unless the separation along the line of sight is very large.
3) As suggested by the relative line-of-sight geometry of NGC 6670E, the C concentration, and NGC 6670W, 
the galaxies are approaching each other again.  This can only happen when the galaxies are gravitationally 
bound.

In addition, the formula above for $T/|U|$ can be used to better constrain the galaxies' transverse velocities 
perpendicular to the line of sight.  The relative velocity ($\triangle V$) of the galaxies at the previous 
pericenter can be estimated by using the lower limit of the pericentric distance (25 kpc) for $R$ and assuming
$T/|U|=1$.  The estimated relative velocity is 210 km $\rm s^{-1}$.  This is slightly larger than the present
radial velocity difference of the galaxies (160 km $\rm s^{-1}$).  Because the galaxies' relative velocity
reached its maximum at the pericenter, the 210 km $\rm s^{-1}$ estimated above must be an upper limit
for the present relative velocity.  Therefore, the observed radial velocity difference suggests that the transverse
velocity must be small compared to the radial velocity.  This is consistent with the discussion at the beginning of 
this section. 

In conclusion, the galaxies in NGC 6670 are likely in gravitationally bound prograde-prograde orbits with
an orbital plane nearly edge-on along the E-W direction.  The galaxies have already passed through a
pericenter at least once.  The pericentric distance of the first close encounter is greater than the diameter
of each stellar disk, therefore no strong stellar tails were generated during that encounter.  We summarize
the discussed orbital geometry in Figure~\ref{cartoon}.  The above general picture of the interaction in 
NGC 6670 is self-consistent, but it should be verified by numerical simulations.

\subsubsection{\ion{H}{1} Concentration between the Stellar Disks}\label{c-concent}

Our \ion{H}{1} observations reveal a massive atomic gas concentration (the C concentration) in the overlap 
region between the two stellar disks.  The C concentration has an \ion{H}{1} mass of  
$\sim2\times10^9 M_{\sun}$.  In some systems of intermediate merging stage such as Arp 244 
\citep{stanford90,lo2000,gao01}, Arp 299 \citep{sargent91,aalto97,casoli99}, and NGC 6090
\citep{gaoetal99,bryant99}, CO images reveal massive molecular gas reservoirs ($>10^9 M_{\sun}$) in the 
disk overlap regions which seem to be the most active star forming areas and could dominate the overall 
starburst activity \citep{wynn91,zhou93,mirabel98}.

As indicated in \S~\ref{h1motion}, the \ion{H}{1} velocities of the E and W concentrations match the
CO velocities of the eastern side of NGC 6670E and the western side of NGC 6670W, respectively.
This suggests that these two \ion{H}{1} concentrations are remnants of the outer \ion{H}{1} disks of the
galaxies and still have velocities similar to the rotation velocities of the inner disks.  In contrast,
Figure~\ref{pvplots} shows that the C concentration has a velocity range between the CO velocities of
the western side of NGC 6670E and the eastern side of NGC 6670W.  
The spatial location and anomalous velocity of the C concentration could be explained by either direct
cloud-cloud collisions of \ion{H}{1} between the prograde-prograde disks, or a superposition of two 
independent \ion{H}{1} components whose velocities are slowed down by the tidal potential 
(\S~\ref{h1motion}).  We note that, even for the second case, cloud-cloud collisions may still take place in
the future because the galaxies are now approaching each other.   \citet{jog92} showed that \ion{H}{1} 
cloud-cloud collisions in colliding galaxies can produce a hot ionized over-pressure remnant gas to
compress the GMCs and lead to starbursts in off-nuclear regions.  The observed concentration of 
\ion{H}{1} between the two disks of NGC 6670 may provide support to the starburst mechanism 
proposed by \citet{jog92}.  

Numerical simulations do not generally produce gas concentrations in the disk overlap regions during the early
or intermediate stages of merging (Arp 244 for example, \citet{mihos93}).  Our \ion{H}{1} observations of
NGC 6670 provide a possible evolutionary link to the intermediate-stage mergers with molecular gas
concentration in the disk overlap regions, like in Arp 244, Arp 299 and NGC 6090.  The C concentration in
NGC 6670 has an \ion{H}{1} gas mass comparable to the molecular concentrations found in the disk overlap
regions of other LIRGs.  Knowing that the C concentration is massive and the galaxies are still approaching
each other, we expect that during the future interaction, more \ion{H}{1} gas (and even molecular gas) will 
flow into this region and get trapped.  
After the gas density increases, a conversion of \ion{H}{1} into $\rm H_2$ may occur.  It is possible that 
the \ion{H}{1} C concentration is the progenitor of the molecular concentration in the disk overlap regions
observed in some late stage LIRGs.

\subsection{Is NGC 6670 Undergoing Starbursts?}
Both galaxies in NGC 6670 have signatures which are anticipated by numerical simulations to be important for 
fueling the nuclear starbursts.  Our BIMA CO observations reveal that the molecular gas in NGC 6670E
is concentrated toward its nuclear region.  A molecular bar is suggested in the center of NGC 6670W
via the CO $p$-$v$ diagram (Fig.~\ref{pvplots}b), which can efficiently drive gas into the nuclear region
\citep{mihos96}.    The IR luminosity of NGC 6670, of $3.8\times10^{11} L_{\sun}$ implies a high star 
formation rate of many tens of solar masses per year.  Furthermore, the total molecular gas mass revealed by 
our observations is also high ($3.6\times10^{10} M_{\sun}$).  However, it is possible that the
GMCs of NGC 6670 are forming stars with an efficiency similar to the GMCs in the Galactic disk.
If this is the case, the high infrared luminosity of NGC 6670 would be merely due to the large reservoirs of
molecular gas, similar to the nonstarburst LIRG Arp 302 \citep{lo97}.  Therefore, it is important to determine
whether the star formation activity in the large molecular reservoirs of NGC 6670 are really elevated to
starburst levels.

\subsubsection{The Star Formation Efficiency}\label{SFE}
There are several ways to probe starbursts.  If the infrared luminosity is entirely due to dust heated by recent
star formation, the star formation rate (SFR) would be directly proportional to the infrared luminosity.
Various conversions between infrared luminosity and star formation rate have been suggested,
from SFR ($M_{\sun}$ $\rm yr^{-1}$) $\sim 77$ to 300 $\times 10^{-12}$ $L_{\rm IR}$($L_{\sun}$)
\citep{scoville83,gallagher86,hunter86,inoue2000}.  Thus the global star formation rate for NGC 6670 is 
$sim30$ to 120 $M_{\sun}$ $\rm yr^{-1}$.  This implies a molecular gas depletion time between $10^8$ 
and $10^9$ yr, comparable with the orbital time scale of NGC 6670 and the typical galaxy-galaxy 
merging time scale for gas-rich mergers demonstrated in most simulations.

The ratio $\rm SFE\it \equiv L_{\rm IR}/M_{\rm H_2}$ is often used to indicate the star formation
efficiency (SFR per unit molecular gas mass).  The mean SFE for the GMCs in the Milky
Way's disk is $\sim 4$ $L_{\sun}/M_{\sun}$ \citep{scoville89}.  For nearby starburst galaxies,
the value of SFE is $\sim10$ $L_{\sun}/M_{\sun}$ \citep{sanders96}.  For nuclear starburst regions in
LIRGs, the SFE ranges from 20 to greater than 100 $L_{\sun}/M_{\sun}$ \citep{scoville91,bryant99}. 
The global SFE for NGC 6670 is $\sim10$ $L_{\sun}/M_{\sun}$.

In order to investigate the location and spatial extent of the starburst, the correlation between the FIR and
radio continuum flux densities (see, e.g., \citet{condon91a}) can be applied to the individual galaxies of NGC
6670.  This should be a good approximation for NGC 6670 because we have shown that the system as a
whole obeys the FIR--radio correlation and that the radio continuum emission of NGC
6670 is not from AGN activity (\S~\ref{20cmresult}).  After scaling the infrared luminosity with the 20 cm
flux density (Table~\ref{tab2}), we find an SFE of 12.7 $L_{\sun}/M_{\sun}$ for NGC 6670E and 8.8
$L_{\sun}/M_{\sun}$ for NGC 6670W.  The SFE of NGC 6670E is significantly higher than that of the
GMCs in the Galactic disk but is not as high as the SFE of nuclear starbursts in LIRGs.

In normal galaxies, the FIR--radio correlation appears to hold locally within individual galaxies and the FIR and 
radio morphologies are similar \citep{bicay88,condon92,marsh95,lu96}.  This is also the case in NGC 6670, as 
the radio continuum and mid-infrared morphology observed by ISO are similar \citep{hwang99}.  If we assume 
the distribution of FIR emission is similar to that of the radio continuum, the strong radio emission from the nuclei 
of NGC 6670 would imply that the nuclear regions are the sites of current star formation activity.  We plot 
one-dimensional profiles of 20 cm flux density and CO integrated intensity along the major axes of the two disks 
in Figure~\ref{profiles}.  It shows that the ratio of 20 cm flux density to CO integrated intensity is significantly 
higher in the center of NGC 6670E compared to other regions in NGC 6670E or NGC 6670W.  Thus, the SFE 
in the center of NGC 6670E is higher than the mean SFE of NGC 6670E or the whole NGC 6670 system.

To quantify the SFE in the nuclear regions, we use two methods to estimate the 20 cm and CO flux from the
central regions.  1) We fit the nuclear components of both galaxies in the high resolution CO integrated 
intensity map and the 20 cm map with 2D Gaussians.  Using this method, we obtain SFEs of 17 and 10
$L_{\sun}/M_{\sun}$ for the nuclear regions of NGC 6670E and NGC 6670W, respectively.  2) We directly 
sum the pixels within the inner 4 kpc ($\sim7\arcsec$) regions of both galaxies in the 20 cm map and the high
resolution CO intensity map to obtain the flux.  The resulting SFE is 18 $L_{\sun}/M_{\sun}$ for the nucleus of
NGC 6670E and 11 for the nucleus of NGC 6670W.  Table~\ref{tab3} summarizes the parameters given by the
two methods described above.

The SFE estimated above is likely to be a lower limit, since the molecular mass may be over-estimated by 
using a CO-to-$\rm H_2$ conversion factor $X$ derived from the Milky Way disk (\S~\ref{codist}).  
We conclude that the nucleus of NGC 6670E is a starburst region and its star formation efficiency is at least 
four times higher than that of the GMCs in the Milky Way disk.

\subsubsection{Other Indicators}
We can calculate the peak CO brightness temperature ($T_b$) using the channel maps.   In order to
minimize the effect of beam dilution, we use the highest resolution maps which include B array data only.
For a $1.8\arcsec$ ($\sim1.1$ kpc) beam, the highest CO brightness in NGC 6670E appears in its center,
in the channel of $v=8740$ km $\rm s^{-1}$.  This brightest component is unresolved and has a peak
brightness temperature of 4.5 K.  In NGC 6670W, several peaks with $T_b$ of 3 to 3.5 K appear in the
disk and we cannot distinguish an isolated peak in its center.  In the starburst galaxy M82, the molecular 
clouds smaller than 100 pc have $T_b$ ranging from 13 to 34 K \citep{shen95}.   At the same time, the 
highest $T_b$ observed by \citet{rickard77} with a $65\arcsec$ single dish beam ($\sim 1$ kpc) on M82
is 0.6 K.  The single dish beam size of \citet{rickard77} is comparable with the synthesis beam used here
for NGC 6670.  Compared with those M82 data, the brightness temperature in the center of NGC 6670E 
and NGC 6670W is unusually high.  Although our resolution is insufficient to resolve individual clouds, 
this still suggests that the molecular clouds in the centers of both NGC 6670E and NGC 6670W are very 
bright and may have properties similar to the clouds in the starburst regions of M82.

We estimate the average infrared surface brightness of the nuclear regions using the FIR--radio correlation
to scale the IR emission with 20 cm radio continuum flux.  Assuming that all the nuclear 20 cm flux given
by the Gaussian fit comes from the area within one full-width-half-maximum (FWHM, see Table~\ref{tab3}), 
we estimate the mean 
IR surface brightness to be $2.4\times10^4$ and $2.0\times10^4$ $L_{\sun}$ $\rm pc^{-2}$ for the 
$\sim 2$ kpc nuclear regions of NGC 6670E and NGC 6670W, respectively.  The typical IR surface 
brightness of resolved starburst regions of few hundred pc sizes in M82, NGC 253 and other LIRGs is 
$\sim 10^5$ $L_{\sun}$ $\rm pc^{-2}$ (see \citet{lo87} for a summary).  In the Orion molecular cloud 
(OMC 1) near the Trapezium, the mean radiation field strength is $\sim 2\times 10^5$ $L_{\sun}$ 
$\rm pc^{-2}$ \citep{tielens85}.  The mean values for the inner parts of NGC 6670 are roughly 10 times 
smaller.  Because our 20 cm beam size ($\sim 2$ kpc) is several times larger than the 
sizes of other resolved starbursts ($<1$ kpc), the FWHMs given by the fitting are clearly upper limits for the 
sizes of the nuclear star forming regions in NGC 6670.  Therefore, the mean value of $\sim 2 \times 10^4$
$L_{\sun}$ $\rm pc^{-2}$ obtained over $\sim 2$ kpc scale is high and the IR surface brightness in the 
nuclear regions of NGC 6670 may be comparable with those in other starburst regions.

The measured CO brightness temperature and IR surface brightness are both largely limited by the resolution
and only represent lower bounds to the actual values.  However, they qualitatively provide tests on the
molecular gas properties and the radiation flux coming from young massive stars.  They both indicate that the
nuclei of NGC 6670E and  NGC 6670W are undergoing higher than normal levels of star formation, perhaps
even at the level of starbursts.  This is consistent with the star formation efficiency estimated in \S~\ref{SFE}.

\subsubsection{What Produced the Elevated SFE?}
In NGC 6670, all the indicators discussed above suggest that the nuclei have properties similar to other
starburst regions.  IN NGC 6670E, the observed accumulation of molecular gas in the central $\sim 2$ 
kpc region may be responsible for the elevated star formation activity.  The detailed accumulation 
mechanism, however, is not clear.  The optical, \ion{H}{1}, and CO observations all show distorted 
morphologies.  The kinematics of the atomic and molecular gas are also disturbed.  It appears that the tidal
interaction has disturbed the gas and stars in the disk of NGC 6670E.  On the other hand, in the center of 
NGC 6670W, much of the molecular gas appears to be in a central molecular bar but the star formation 
efficiency is not as high as in the nucleus of NGC 6670E.  It is unclear in our observations why the intense 
bar driven gas inflow predicted by the simulations does not take place in NGC 6670W.  With that NGC 6670E
has higher kinematic disturbance and star formation efficiency than those of NGC 6670W, our observations 
appear to support the hypothesis that the starburst is fueled by the gas inflow due to the tidal perturbation 
produced by the interaction of the galaxies.

In recent simulations of \citet{mihos96}, the starburst history in merging disk galaxies sensitively depends on
the structure of the progenitor galaxies.  Collisions between disk/halo galaxies produce gas inflows shortly
after the first close passage and thus produce starbursts while the galaxies are still widely separated.  On the
other hand, collisions between disk/bulge/halo galaxies can only produce strong gas inflows and starbursts
in the final stage of the mergers.  Our observations of NGC 6670 show that starbursts can be triggered when
the galaxies are still widely separated, which according to \citet{mihos96} would imply that the NGC 6670 
galaxies fit the disk/halo type.   However, bulges of the two galaxies clearly exist in the NIR images
(Fig~\ref{opticals}).   Our observations do not appear to well agree with the model.   Similarly, there is 
another ULIRG found to be in an early stage of merging and also have a bulge (IRAS 01521+5224, 
\citep{murphy2000}).   To further confirm the structural dependence predicted by the simulations, 
observations of the structures of the NGC 6670 galaxies as well as other early stage LIRGs are needed.

\section{Summary}
1. Very long \ion{H}{1} tidal tails in NGC 6670, a pair of two overlapping edge-on galaxies, are revealed by 
our VLA observations, which proves that the two galaxies are interacting gravitationally.  We have compared 
the morphology and motion of \ion{H}{1} in NGC 6670 with the studies of \citet{toomre72} to reconstruct a 
self-consistent orbital geometry for this interacting system.  The interaction is prograde-prograde with an 
orbital plane nearly edge-on in the E-W direction, with the galaxy spin vectors and orbital vectors essentially 
all parallel to each other.  The galaxies appear to be in an early stage of merging, and have already passed a
pericenter roughly $\sim 4 \times 10^8$ yr ago.  Their pericentric distance of the previous encounter is at least 
25 kpc.  The galaxies are most likely gravitationally bound and are probably approaching another close 
encounter.

2. The total \ion{H}{1} mass in the NGC 6670 system is $1.6\times10^{10} M_{\sun}$.  Unlike other
intermediate or late stage mergers, only 25\% of NGC 6670's \ion{H}{1} gas is distributed within the tails.
Most of the \ion{H}{1} in this system was found near the two stellar disks. The \ion{H}{1} disks of the 
galaxies are interacting with each other and have slower rotational velocities than those of the CO disks.
The global \ion{H}{1} kinematics appears to be dominated by the rotation of the more massive galaxy 
NGC 6670W.  The \ion{H}{1} near NGC 6670W appears to be in a rotating ring-like structure.
The \ion{H}{1} disk of NGC 6670E has been nearly destroyed by the interaction and the \ion{H}{1} 
motion near NGC 6670E appears to be highly perturbed.    

3. Three apparent \ion{H}{1} gas concentrations have been identified at the edges of the stellar disks.  
The most massive one, the C concentration, is located in the region between the stellar disks and has a
collective \ion{H}{1} mass of $2\times10^9 M_{\sun}$.  The C concentration might be a true concentration
formed due to collisions of \ion{H}{1} clouds, but a superposition of two spatially separated components
is also possible.  We need higher resolution \ion{H}{1} data to resolve this ambiguity.
This \ion{H}{1} concentration may be accumulating more \ion{H}{1} gas and molecular
gas during the future interaction of the galaxies, to become the next site of starburst in this system.

4. BIMA CO(1--0) observations reveal large molecular gas reservoirs in NGC 6670.  The molecular gas
mass is $1.1\times10^{10} M_{\sun}$ for NGC 6670E and $2.5\times10^{10} M_{\sun}$ for NGC
6670W.  The molecular clouds in both galaxies are concentrated toward the nuclear regions, especially in
NGC 6670E.  A large portion of the CO flux in NGC 6670E and in NGC 6670W comes from the inner 4
kpc regions of the galaxies.  The estimated face-on molecular gas column density of each galaxy is $\sim10^3$
$M_{\sun}$ $\rm pc^{-2}$, larger than the value of the Galactic center (40--300 $M_{\sun}$ $\rm pc^{-2}$)
but smaller than other LIRGs ($>10^4$ $M_{\sun}$ $\rm pc^{-2}$) in advanced merging stages.  The rotation
curve of NGC 6670W suggests that there may exist a molecular bar in the inner disk.  Both molecular disks
exhibit normal rotational signatures.  The kinematics and morphology of the molecular disks in the two 
galaxies appear less disturbed than the \ion{H}{1} disks, probably due to their greater resistance to tidal 
disruption.

5. Our VLA 20 cm radio continuum imaging reveals the distribution of the radio continuum emission from
the galaxies.  The 20 cm flux density and FIR flux density of NGC 6670 appear to obey the FIR--radio
correlation for normal galaxies so AGN activity is ruled out in this system.  In each galaxy, more than 50\%
of the radio continuum flux comes from the inner 4 kpc region.  This suggests active star formation
in the nuclear regions of the galaxies.

6. The FIR--radio correlation was applied locally to the galaxies' nuclear regions.  We estimate the star
formation efficiency of the central regions of both galaxies by scaling the FIR luminosity by the radio
continuum flux.  The nuclei of NGC 6670E and NGC 6670W are estimated to have the values of SFE
$\sim18$ and $\sim 11$ $L_{\sun}/M_{\sun}$, respectively.   Those values are significantly larger than
typical SFEs of GMCs in the Galactic disk ($\sim4$ $L_{\sun}/M_{\sun}$) and are nearly as high as that
found in starbursts and in other LIRGs (20 to $>$100 $L_{\sun}/M_{\sun}$).  The star formation efficiency 
and other indicators of starbursts such as high CO brightness temperature and FIR surface brightness all 
suggest that the central regions of the two galaxies have properties similar to starburst regions in other 
galaxies.

\acknowledgments
We thank D.-C. Kim and D.B. Sanders for the use of the $K^{\prime}$-band image of NGC 6670.
We would like to thank the referee of ApJ Letters, Pierre-Alan Duc, for his invaluable comments 
and suggestions that have greatly helped to improve this paper.  We are also very grateful to the referee
of AJ, Tom Murphy, for his kindly providing an alternative interpretation to our \ion{H}{1} data and the 
very useful comments.  W.H. Wang and K.Y. Lo acknowledge partial support from the Academia Sinica
and the National Science Council in Taiwan.  K.Y. Lo, Yu Gao and Robert Gruendl acknowledge partial 
support from the Laboratory of Astronomical Imaging which is funded by NSF grant AST 96-13999 and 
by the University of Illinois.  This study of NGC 6670 was started when W.H. Wang was a master degree 
student in the Institute of Astronomy in National Central University.  W.H. Wang thanks the support from
NCU.  Yu Gao is grateful to the ASIAA for its hospitality during a short visit when the writing of this paper 
was initiated.

\appendix
\section{CGCG 301-032}\label{cgcg301032}
We detect \ion{H}{1} emission at the velocity range between 8630 and 8870 km $\rm s^{-1}$ from
the galaxy CGCG 301-032 (IRASF 18335+5949), which is $\sim 5\arcmin$ away in the southeast of
NGC 6670.  The low resolution \ion{H}{1} integrated intensity map and velocity map of CGCG
301-032 are plotted in Figure~\ref{cgcgmoms}.  There is no \ion{H}{1} detected between NGC 6670
and CGCG 301-032 under the current detection limit.  The \ion{H}{1} integrated flux of CGCG 301-032 
obtained from the low resolution channel maps is 1.6 Jy km $\rm s^{-1}$ and indicates the total 
\ion{H}{1} gas mass to be $5.5\times10^9 M_{\sun}$.  The extent of the HI gas is about 30 kpc.  
The galaxy's \ion{H}{1} velocity gradient along NE-SW direction 
shows that the \ion{H}{1} of CGCG 301-032 is in a rotating 
disk.  The large amount of atomic gas and the \ion{H}{1} rotating feature suggest that CGCG 301-032 
is a relatively small (12 kpc in optical extent)  early-type spiral galaxy.  However, the spiral arms are not 
apparent in the DSS image and our $J$-band image.  This galaxy is also detected in our 20 cm radio
continuum map.  The total flux density of CGCG 301-032 at 20 cm is 5.9 mJy.

\clearpage

\figcaption[]{A comparison of the optical to near-infrared images of NGC 6670  at 
\protect$B, V, R, I, J,$ and \protect$K^{\protect\prime}$-bands.\label{opticals}}

\figcaption[]{Robustly weighted \protect\ion{H}{1} moment maps at
\protect$18\protect\arcsec\protect\times15\protect\farcs6$ resolution of NGC 6670.
(a) The \protect\ion{H}{1} integrated intensity contours are overlaid on the \protect$J$-band grey-scale image.  The
contour levels are 1.5, 6, 10, 15, 20, 25, 30, 40, 50, and 60 times 8.5 mJy \protect$\protect\rm beam^{-1}$ km
\protect$\protect\rm s^{-1}$ which corresponds to an \protect\ion{H}{1} column density of \protect$3.3\protect\times10^{19}\protect\rm cm^{-2}$.
Several \protect\ion{H}{1} components mentioned in the text
are labeled.  (b) The \protect\ion{H}{1} intensity-weighted radial velocity contours on \protect\ion{H}{1} integrated
intensity grey-scale image.  The velocity difference between all adjacent contours is 40 km \protect$\protect\rm s^{-1}$.
\label{h1moms}}

\figcaption[]{Uniformly weighted \protect\ion{H}{1} integrated intensity at \protect$11\protect\farcs4$ resolution 
of NGC 6670.  The \protect\ion{H}{1} intensity is plotted with grey contours and the CO(1--0) integrated intensity 
is also plotted with black contours for comparison.  The \protect\ion{H}{1} contour levels are 2, 4, 6, ..., 24 times 
18 mJy \protect$\protect\rm beam^{-1}$ km \protect$\protect\rm s^{-1}$ which corresponds to an \protect\ion{H}{1} column density of
\protect$1.5\protect\times10^{20}\protect\rm cm^{-2}$.  The underlying grey-scale picture is the \protect$K^{\protect\prime}$-band image of 
D.-C Kim and D. B. Sanders.  \label{h1unifmom0}}

\figcaption[]{\protect\ion{H}{1} 
spectra of the whole NGC 6670 system, the W tail, the NE tail, and the C concentration. 
The spectrum of the C concentration is obtained from the high resolution cube.  Others are
obtained from the low resolution cube. \label{h1spect}}

\figcaption[]{\protect\ion{H}{1} channel map contours at $18\protect\arcsec\protect\times15\protect\farcs6$ resolution, 
overlaid on the \protect$J$-band grey-scale image.  Contours start at 0.69 mJy \protect$\protect\rm beam^{-1}$ (\protect$2\protect\sigma$) and 
increase by steps of 0.69 mJy \protect$\protect\rm beam^{-1}$.  Plotted in the upper-left corner of each sub-plot is the 
central heliocentric velocity of the channel. \label{h1chmap}}

\figcaption[]{\protect\ion{H}{1} and CO position-velocity 
(\protect$p$-\protect$v$) diagrams: (a) along the optical disk of NGC
6670E (P.A.\protect$=120\protect\arcdeg$), (b) along the optical disk of NGC 6670W (P.A.\protect$=70\protect\arcdeg$), (c) along the
line connecting the two nuclei of NGC 6670E and NGC 6670W (P.A.\protect$=76\protect\arcdeg$).  The \protect\ion{H}{1}
$p$-$v$ diagrams obtained from the low resolution cube are plotted with thin lines filled with grey colors.
The \protect\ion{H}{1} contours shown in all the three plots represent \protect\ion{H}{1} brightness starting at 0.7 mJy
\protect$\protect\rm beam^{-1}$ (\protect$2\protect\sigma$) and increasing by steps of 0.7 mJy \protect$\protect\rm beam^{-1}$.  The grey color is 
smoothed \protect\ion{H}{1} brightness.  In the three plots, CO \protect$p$-\protect$v$ diagrams made along the same position 
angles as \protect\ion{H}{1} are plotted with thick lines. The CO \protect$p$-\protect$v$ diagrams were obtained from the low 
resolution cube.  The CO contour levels are 1, 2, 3,..., 7 times 26 mJy \protect$\protect\rm beam^{-1}$ (\protect$1.8\protect\sigma$).  
In (a), the origin of the position axis is at the center of NGC 6670E.  In (b) and (c), the origins are at the 
center of NGC 6670W.  The cross signs in each plot represent the \protect\ion{H}{1} (thin lines) and CO (thick 
lines) resolutions in both velocity and spatial axes. The velocity resolution is 21.8 km \protect$\protect\rm s^{-1}$ for 
\protect\ion{H}{1} and is 20 km \protect$\protect\rm s^{-1}$ for CO. The spatial resolutions for \protect\ion{H}{1} and CO are determined 
from the beam sizes along the direction of the slices and change slightly in each plot. The central positions of 
the \protect\ion{H}{1} W concentration, the C concentration, and the E concentration are indicated with horizontal 
grey lines in (a) and (b).  The central velocities of NGC 6670E and NGC 6670W are indicated with vertical 
grey lines.  The scales in both axes are different in all the three plots. \label{pvplots}}

\figcaption[]{CO(1--0) channel map contours at \protect$2\protect\farcs7\protect\times2\protect\farcs1$ resolution overlaid 
on the grey-scale \protect$K^{\protect\prime}$-band image.  The contour levels are 1, 2, 3, 4, and 5 times 26
mJy \protect$\protect\rm beam^{-1}$ (\protect$=2\protect\sigma=0.4$ K). The central velocity of each channel is labeled in the upper-right
corner of every sub-plot. \label{cochmap}}

\figcaption[]{CO(1--0) integrated intensity maps.  (a) Intensity map at \protect$5\protect\farcs4\protect\times4.\protect\arcsec6$ 
resolution.  The CO integrated intensity is plotted as contours overlaid on the \protect$K^{\protect\prime}$-band image 
provided by D.-C. Kim and D.B. Sanders.  Contour levels are 1, 2, 3, 4, 6, 8, 10, 12, 14, 18, 22, and 26 times 2.2 Jy
\protect$\protect\rm beam^{-1}$ km \protect$\protect\rm s^{-1}$.  (b) Intensity map at \protect$2\protect\farcs7\protect\times2\protect\farcs1$
resolution.  The CO intensity contours are overlaid on the \protect$B$-band grey-scale image.  The contour levels are 1, 2, 3, 4, 6, 8, and
10 times 2.4 Jy \protect$\protect\rm beam^{-1}$ km \protect$\protect\rm s^{-1}$ 
(\protect$=1.1\protect\times10^{22}$ \protect$\protect\rm cm^{-2}$). \label{comoms}}

\figcaption[]{CO(1--0) spectra of NGC 6670E and NGC 6670W obtained from 
the higher resolution channel maps.  \label{cospect}}

\figcaption[]{CO radial velocity contours derived from the low resolution intensity-weighted 
radial velocity map, overlaid on the grey-scale CO integrated intensity. The contours are plotted every 30 km 
\protect$\protect\rm s^{-1}$. \label{comom1}}

\figcaption[]{20 cm radio continuum contours of NGC 6670 overlaid on (a) grey-scale of the 
\protect$B$-band optical image and (b) grey-scale of the high resolution CO(1--0) integrated intensity.  The contours in
each sub-plot are 1, 2, 4, 8, 16, 32, 64, and 128 times 0.1 mJy \protect$\protect\rm beam^{-1}$ (\protect$=2\protect\sigma=5.4$ K).  The 
black circles represent the 20 cm beam and the grey ellipse represents the CO beam.
\label{20cm}}

\figcaption[]{CO(1--0) integrated intensity and 20 cm radio continuum 
intensity profiles along the major axes of NGC 6670W (P.A. \protect$=70\protect\arcdeg$) and NGC 6670E (P.A.
\protect$=120\protect\arcdeg$).  The CO intensity is plotted with solid lines and the 20 cm intensity is plotted with dashed 
lines.  The 20 cm intensity is labeled in the left-hand-side of each plot.  The CO intensity is labeled in the 
right-hand-side of each plot.  In each plot, the zero of the x-axis is located in the center of the galaxy.
\label{profiles}}
 
\figcaption[]{A cartoon depiction of a possible orbital geometry of NGC 6670.  
Two major ideas of this cartoon discussed in \protect\S~\protect\ref{interaction2} are 1) the galactic planes and the 
orbital plane are all viewed nearly edge-on 2) NGC 6670E is in front of NGC 6670W along the line 
of sight.  The radial systemic velocities of the galaxies (\protect$V_{sys}$) are shifted so that NGC 6670W has 
zero radial velocity.   The rotational velocities (\protect$V_{rot}$) are determined from the CO velocity filed.
We note that the orbital geometry presented by this cartoon is consistent with the observations but is 
not necessarily the unique possibility.  Numerical simulations are needed to verify this 
orbital geometry.  \label{cartoon}}

\figcaption[]{Contours of the low resolution \protect\ion{H}{1} intensity and 
radial velocity of CGCG 301-032.
(a) Integrated \protect\ion{H}{1} intensity contours at 1.5, 6, 10, 15, 20, 25, 30, 40, and 50 times 8.5 mJy
\protect$\protect\rm beam^{-1}$ km \protect$\protect\rm s^{-1}$ (\protect$3.3\protect\times10^{19}\protect\rm cm^{-2}$), 
overlaid on a \protect$J$-band image.
(b) radial velocity contours overlaid on the grey-scale \protect\ion{H}{1} integrated intensity.  The contours are
plotted every 25 km \protect$\protect\rm s^{-1}$.  \label{cgcgmoms}}

\clearpage
\begin{deluxetable}{lccccc}
\tablecaption{Observed \ion{H}{1} Properties of NGC 6670 \label{tab1}}
\tablehead{\colhead{HI Component}  &
\colhead{$S_{\rm 21cm}$\tablenotemark{a}}  &
\colhead{$M_{\rm HI}$}  &
\colhead{$<$$\sigma_v$$>$\tablenotemark{b}}  &
\colhead{$n_{\rm HI, peak}$/$<$$n_{\rm HI}$$>$ \tablenotemark{c}}  &
\colhead{$l_{max} \times l_{min}$\tablenotemark{d}} \\
\colhead{}  &
\colhead{(Jy km $\rm s^{-1}$)}  &
\colhead{($10^9M_{\sun}$)}  &
\colhead{(km $\rm s^{-1}$)}  &
\colhead{($10^{20}$ atoms $\rm cm^{-2}$)}  &
\colhead{($\rm kpc \times kpc$)}
}
\startdata
NGC 6670 (total)\tablenotemark{1} &  4.83  &  16.4  & \nodata &  \nodata       &  $110\times30$ \\
W tail\tablenotemark{1}                  &  0.97  &  3.3   &  16.5    &  5.1 / 2.8      &  $60\times20$  \\
NE tail\tablenotemark{1}                 &  0.24  &  0.8   &  10.7    &  3.9 / 2.5      &  $15\times15$  \\
C concentration\tablenotemark{2}   &  0.57  &  2.0   &   34.5    &  $>24.0$ / 14.0  &  $14\times13$  \\
W concentration\tablenotemark{2}  &  0.40  &  1.4   &   34.9    &  $>38.0$ /         &  unresolved      \\
\enddata
\tablenotetext{1}{Properties of the extended \ion{H}{1} components are derived from the low resolution
(robustly weighted) maps.}
\tablenotetext{2}{Properties of the concentrated components are derived from the high resolution (uniformly
weighted) maps.}
\tablenotetext{a}{Integrated \ion{H}{1} line flux obtained from the channel maps.}
\tablenotetext{b}{\ion{H}{1} mean velocity dispersion derived from the intensity-weighted velocity dispersion
maps.}
\tablenotetext{c}{Peak and averaged \ion{H}{1} column density derived from the integrated intensity maps.}
\tablenotetext{d}{Maximum and minimum dimensions measured from the first outer contours in the 
integrated intensity maps.}
\end{deluxetable}

\clearpage
\begin{deluxetable}{lcc}
\tablecaption{Global CO and Radio Continuum Properties of NGC 6670 \label{tab2}}
\tablehead{\colhead{Derived Parameter}  &
\colhead{NGC 6670E}  &
\colhead{NGC 6670W}}
\startdata
$S_{\rm CO}$\tablenotemark{a} ~~(Jy km $\rm s^{-1}$)  &  65.4  &  149.7  \\
$M_{\rm H_2}$ ~~($10^{10} M_{\sun}$)  &  1.1  &  2.5  \\
$n_{\rm H_2, peak}$\tablenotemark{b}~~($10^{22}$ molecules $\rm cm^{-2}$)  &  8.0  &  11.8  \\
$N_{\rm H_2, peak}$\tablenotemark{b}~~($10^3 M_{\sun}$ $\rm pc^{-2}$) &  1.3  &  1.9  \\
$S_{\rm 20cm}$~~(mJy)  &  22.3  &  35.6  \\
$L_{\rm IR}$\tablenotemark{c}~~($10^{11} L_{\sun}$)  &  1.4  &  2.2  \\
SFE\tablenotemark{d}~~($L_{\sun}/M_{\sun}$)  &  12.7  &  8.8  \\
\enddata
\tablenotetext{a}{CO integrated flux obtained from the high resolution channel maps.}
\tablenotetext{b}{Peak edge-on $\rm H_2$ column density derived from the high resolution intensity map.}
\tablenotetext{c}{IR luminosity calculated by scaling the total IR luminosity with the 20 cm flux.
See \S~\ref{SFE}.}
\tablenotetext{d}{Star formation efficiency $\equiv L_{\rm IR}/M_{\rm H_2}$.  See \S~\ref{SFE}.}
\end{deluxetable}

\clearpage
\begin{deluxetable}{lcc}
\tablecaption{Properties of the Nuclear Regions of NGC 6670 \label{tab3}}
\tablehead{\colhead{Derived Parameter}&\colhead{Center of NGC 6670E}&\colhead{Center of NGC 6670W}}
\startdata
By Gaussian Fit &  &  \\
~~~~$L_{\rm IR}$~~($10^{10} L_{\sun}$)  &  9.5  &  12.6 \\
~~~~$\Sigma_{\rm IR}$\tablenotemark{a}~~($10^4 L_{\sun}$ $\rm pc^2$)  & 2.4 & 2.0 \\
~~~~$M_{\rm H_2}$~~($10^9 M_{\sun}$)  &  5.6  &  12.5  \\
~~~~SFE~~($L_{\sun}/M_{\sun}$)  &  17  &  10  \\
~~~~$l_{maj}\times l_{min}$\tablenotemark{b}~~($\rm kpc \times kpc$)  &  $2.5\times2.0$  &  $3.4\times2.3$ \\
By Direct Measuring &  &  \\
~~~~$L_{\rm IR}$~~($10^{10} L_{\sun}$)  &  9.0  &  10.5  \\
~~~~$M_{\rm H_2}$~~($10^9 M_{\sun}$)  &  5.0  &  9.4  \\
~~~~SFE~~($L_{\sun}/M_{\sun}$)  &  18  &  11  \\
\enddata
\tablenotetext{a}{Averaged infrared surface brightness.}
\tablenotetext{b}{Major and minor axes FWHMs given by the Gaussian fit in the 20 cm map.}
\tablecomments{This table summarizes the properties of the nuclear regions. The CO and 20
cm flux from each galactic nucleus was isolated by 1) applying Gaussian fit to the nuclear regions 2) directly
summing the pixels within the inner 4 kpc $\times$ 4 kpc regions.  Also see the text in \S~\ref{SFE}.}
\end{deluxetable}


\begin{thebibliography}{}
\bibitem[Aalto et al.(1997)]{aalto97}
Aalto, S., Radford, S. J. E., Scoville, N. Z., \& Sargent, A. I. 1997, \apjl, 475, L107
\bibitem[Arp(1966)]{arp66}
Arp, H. 1996, \apjs, 14, 1
\bibitem[Athanassoula(1992)]{athana92}
Athanassoula, E. 1992, \mnras, 259, 345
\bibitem[Barnes \& Hernquist(1991)]{barnes91}
Barnes, J. E., \& Hernquist, L. 1991, \apjl, 370, 65
\bibitem[Bessell, Castelli, \& Plez(1998)]{bessell98}
Bessell, M.S., Castelli, F., \& Plez, B. 1998, \aap, 333, 231
\bibitem[Bicay et al.(1988)]{bicay88}
Bicay, M. D., Helou, G., \& Condon J. J. 1988, \apjl, 338, L53
\bibitem[Broeils \& van Woerden(1994)]{broeils94}
Broeils, A. H., \& van Woerden, H. 1994 \aaps, 107, 129
\bibitem[Briggs(1995)]{briggs95}
Briggs, D. S. 1995, \baas, 187, 112.02
\bibitem[Bryant \& Scoville(1999)]{bryant99}
Bryant, P. M., \& Scoville, N. Z. 1999, \aj, 117, 2632
\bibitem[Bushouse(1987)]{bushouse87}
Bushouse, H. A. 1987, \apj, 320, 49
\bibitem[Bushouse, Werner, \& Lamb(1988)]{bushouse88}
Bushouse, H. A., Werner, M. W., \& Lamb, S. A. 1988, \apj 335, 74
\bibitem[Casoli et al.(1999)]{casoli99}
Casoli, F., Willaime, M.-C., Viallefond, F., \& Gerin, M. 1999, \aap, 346, 663
\bibitem[Condon, Anderson, \& Helou(1991)]{condon91a}
Condon, J. J., Anderson, M. L., Helou, G. 1991, \apj, 376, 95
\bibitem[Condon et al.(1991)]{condon91b}
Condon, J. J., Huang, Z.-P., Yin, Q. F., \& Thuan, T. X. 1991, \apj, 378, 65
\bibitem[Condon(1992)]{condon92}
Condon, J. J. 1992, \araa, 30, 575
\bibitem[Downes \& Solomon(1998)]{downes98}
Downes, D., \& Solomon, P. M. 1998, \apj, 507, 615
\bibitem[Duc, Mirabel, \& Maza(1998)]{duc98}
Duc, P. A., Mirabel, I. F., \& Maza, J. 1998, \aaps, 124, 533
\bibitem[Gallagher \& Hunter(1986)]{gallagher86}
Gallagher, J. S., Hunter, D. A. 1986, in Conf. Proc. of Star Formation in Galaxies, Pasadena 1986,
ed. C. Lonsdale (Washington, DC: GPO), 167
\bibitem[Gao(1996)]{gao96}
Gao, Y. 1996, Ph.D. thesis, State Univ.\ of New York at Stony Brook
\bibitem[Gao et al.(1997)]{gaoetal97}
Gao, Y., Gruendl, R., Lo, K. Y., Hwang, C. Y., Veilleux, S. 1997, in AIP Conf. Ser. 393,
Star Formation Near and Far: Seventh Astrophysics Conference, ed. S. S. Holt \& L. G. Mundy.
(New York: AIP Press), 319
\bibitem[Gao et al.(1999)]{gaoetal99}
Gao, Y., Gruendl, R. A., Hwang, C. Y., Lo, K. Y. 1999, in IAU Symp. 186, Galaxy Interactions at
Low and High Redshift, ed. J. E. Barnes, \& D. B. Sanders, (Dordrecht: Kluwer), 227
\bibitem[Gao \& Solomon(1999)]{gao99}
Gao, Y., \& Solomon, P. M. 1999, \apjl, 512, L99
\bibitem[Gao et al.(2001)]{gao01}
Gao, Y., Lo, K. Y., Lee, S.-W., \& Lee, T.-H. 2001, \apj, 548, 172
\bibitem[Genzel et al.(1998)]{genzel98}
Genzel, R. et al. 1998, \apj, 498, 579
\bibitem[Hibbard \& van Gorkom(1996)]{hibbard96}
Hibbard, J. E., \& van Gorkom, J. H. 1996, \aj, 111, 655
\bibitem[Hibbard \& Yun(1999)]{hibbard99}
Hibbard, J. E., \& Yun, M. S. 1999, \aj, 118, 162
\bibitem[Huchtmeier \& Richter(1989)]{huchtmeier89}
Huchtmeier, W. K., \& Richter, O.-G. 1989, A General Catalog of \ion{H}{1} Observations
of Galaxies (New York: Springer-Verlag)
\bibitem[Hunter et al.(1986)]{hunter86}
Hunter, D. A., Gillett, F. C., Gallagher, J. S., Rice, W. L., \& Low, F. J. 1986, \apj, 303, 171
\bibitem[Hwang et al.(1999)]{hwang99}
Hwang, C.-Y., Lo, K. Y., Gao, Y., Gruendl, R. A., \& Lu, N. Y. 1999, \apjl, 511, L17
\bibitem[Inoue, Hirashita, \& Kamaya(2000)]{inoue2000}
Inoue, A. K., Hirashita, H. \& Kamaya, H. 2000, \pasj, 52, 539
\bibitem[Jog \& Solomon(1992)]{jog92}
Jog, C. J., \& Solomon, P. M. 1992, \apj, 387, 152
\bibitem[Jogee \& Kenny(1996)]{jogee96}
Jogee. S., \& Kenny, J. D. P. 1996, in ASP Conf. Ser. 91, Barred Galaxies,
ed. R. Buta, D. A. Crocker, \& B. G. Elmegreen (San Francisco: ASP), 230
\bibitem[Joseph \& Wright(1985)]{joseph85}
Joseph, R. D., \& Wright, G. S. 1985, \mnras, 214, 87
\bibitem[Kennicutt et al.(1987)]{kennicutt87}
Kennicutt, R. C., Jr., Keel, W. C., van der Hulst, J. M.,
Himmel, E., \& Roettiger, K. A. 1987, \aj, 93, 1011
\bibitem[Kennicutt(1998)]{kennicutt98} Kennicutt, R. C., Jr. 1998, \apj, 498, 541
\bibitem[Kuijken \& Merrifield(1995)]{kuijken95}
Kuijken, K., \& Merrifield, M. R. 1995, \apjl, 443, L13
\bibitem[Lo et al.(1987)]{lo87}
Lo, K. Y., Cheung, K. W., Masson, C. R., Phillips, T. G., Scott, S. L., \& Woody, D. P.
1987, \apj, 312, 574
\bibitem[Lo, Gao, \& Gruendl(1997)]{lo97}
Lo, K. Y., Gao, Y., \& Gruendl, R. A. 1997, \apjl, 475, L103
\bibitem[Lo et al.(2000)]{lo2000}
Lo, K. Y., Hwang, C. Y., Lee, S. W., Kim, D. C., Wang, W. H., Lee, T. H., Gruendl, R., Gao, Y. 2000,
in ASP Conf. Ser. 197, Dynamics of Galaxies: from the Early Universe to the Present, ed.
F. Combes, G. A. Mamon, \& V. Charmandaris (Paris: ASP), 279
\bibitem[Lu et al.(1996)]{lu96}
Lu, N. Y., Helou, G., Tuffs, R., Xu, C., Malhotra, S., Werner, M. W., \& Thronson, H. 1996,
\aap, 315, L153
\bibitem[Maloney \& Black(1988)]{maloney88}
Maloney, P., \& Black, J. H. 1988 \apj, 325, 389
\bibitem[Marsh \& Helou(1995)]{marsh95}
Marsh, K. A., \& Helou, G. 1995, \apj, 445, 599
\bibitem[Martin, Bottinelli, \& Gouguenheim(1991)]{martin91}
Martin, J. M., Bottinelli, L., \& Gouguenheim, L. 1991, \aap, 245, 393
\bibitem[Meixner, Young Owl, \& Leach(1999)]{meixner99}
Meixner, M., Young Owl, R. C. \& Leach, R. 1999, \pasp, 111, 997
\bibitem[Melnick \& Mirabel(1990)]{melnick90}
Melnick, J., \& Mirabel, I. F. 1990, \aap, 231, L19
\bibitem[Mihos, Bothun, \& Richstone(1993)]{mihos93}
Mihos, J. C., Bothun, G. D., \& Richstone, D. O. 1993 \apj, 418, 82
\bibitem[Mihos \& Hernquist(1996)]{mihos96}
Mihos, J. C., \& Hernquist, L. 1996, \apj, 464, 641
\bibitem[Mihos(2001)]{mihos01}
Mihos, J. C. 2001, \apj, in press
\bibitem[Mirabel et al.(1998)]{mirabel98}
Mirabel, I. F. et al. 1998, \aap, 333, L1
\bibitem[Moshir et al.(1990)]{moshir90}
Moshir, M. et al. 1990, IRAS Faint Source Catalogue, version 2.0
\bibitem[Mundell et al.(1995)]{mundell95}
Mundell, C. G., Pedlar, A., Axon, D. J., Meaburn, J., \& Unger, S. W.
1995, \mnras, 277, 641
\bibitem[Murphy et al.(1996)]{murphy96}
Murphy, T. W., Arnus, L., Matthews, K., Soifer, B. T., 
\& Mazzarella, J. M., et al. 1996, \aj, 111, 1025
\bibitem[Murphy(2000)]{murphy2000}
Murphy, T. W. 2000, Ph.D. thesis, California Institute of Technology
\bibitem[Regan \& Gruendl(1995)]{regan95}
Regan, M. W., \& Gruendl, R. A. 1995, in ASP Conf. Proc. Ser. 77,
Astronomical Data Analysis Software and Systems IV,
ed. R. A. Shaw, H. E. Payne, \& J. J. E. Hayes (San Francisco: ASP), 335
\bibitem[Rickard et al.(1977)]{rickard77}
Rickard, L. J., Palmer, P., Morris, M., Turner, B. E., \& Zuckerman, B. 1977, \apj, 213, 673
\bibitem[Rieke et al.(1985)]{rieke85}
Rieke, G. H., Cutri, R., Black, J. H., Kailey, W. F., McAlary, C. W.,
Lebofsky, M. J., \& Elston, R. 1985, \apj, 290, 116
\bibitem[Roberts(1975)]{roberts75}
Roberts, M. S. 1975, in Galaxies and the Universe, ed. A. Sandage, M. Sandage, \& J. Kristian 
(Chicago: University of Chicago Press), 309
\bibitem[Sanders, Solomon, \& Scoville(1984)]{sanders84}
Sanders, D. B., Solomon, P. R., \& Scoville, N. Z. 1984, \apj, 276, 182
\bibitem[Sanders et al.(1988)]{sanders88}
Sanders, D. B., Soifer, B. T., Elias, J. H., Madore, B. F.,
Matthews, K., Neugebauer, G., \& Scoville, N. Z. 1988, \apj, 325, 74
\bibitem[Sanders et al.(1991)]{sanders91}
Sanders, D. B., Scoville, N. Z., \& Soifer, B. T. 1991, \apj, 370, 158
\bibitem[Sanders et al.(1995)]{sanders95}
Sanders, D. B., Egami, E., Lipari, S., Mirabel, I. F., Soifer, B. T. 1995, \aj, 110, 1993
\bibitem[Sanders \& Mirabel(1996)]{sanders96}
Sanders, D. B., \& Mirabel, I. F. 1996, \araa, 34, 749
\bibitem[Sargent \& Scoville(1991)]{sargent91}
Sargent, A. I., \& Scoville, N. Z. 1991, \apjl, 366, L1
\bibitem[Sault, Teuben, \& Wright(1995)]{sault95}
Sault, R. J., Teuben, P. J., \& Wright, M. C. H. 1995, in ASP Conf. Proc. 77,
Astronomical Data Analysis Software and Systems IV,
ed. R. A. Shaw, H. E. Payne, \& J. J. E. Hayes (San Francisco: ASP), 433
\bibitem[Schwarz(1984)]{schwarz84}
Schwarz, M. P. 1984, \mnras, 209, 93
\bibitem[Scoville \& Young(1983)]{scoville83}
Scoville, N., \& Young, J. S. 1983, \apj, 265, 148
\bibitem[Scoville, Sanders, \& Clemens(1986)]{scoville86}
Scoville, N. Z., Sanders, D. B., \& Clemens, D. P. 1986, \apjl, 310, L77
\bibitem[Scoville et al.(1987)]{scoville87}
Scoville, N. Z., Yun, M. S., Clemens, D. P., Sanders, D. B.,
\& Walker, W. H. 1987, \apjs, 63, 821
\bibitem[Scoville \& Good(1989)]{scoville89}
Scoville, N. Z., \& Good, J. 1989, \apj, 339, 140
\bibitem[Scoville et al.(1991)]{scoville91}
Scoville, N. Z. et al.\ 1991, \apjl, 366, L5
\bibitem[Shen \& Lo(1995)]{shen95}
Shen, J., \& Lo, K. Y. 1995, \apjl, 445, L99
\bibitem[Smith et al.(1998)]{smith98}
Smith, H. E., Lonsdale, C. J., Lonsdale, C. J., \& Diamond, P. J. 1998, \apj, 493, L17
\bibitem[Solomon, Downes, \& Radford(1992)]{solomon92}
Solomon, P. M., Downes, D., \& Radford, S. J. E. 1992, \apjl, 387, L55
\bibitem[Solomon et al.(1997)]{solomon97}
Solomon, P. M. et al.\ 1997, \apjs, 478,144
\bibitem[Spinoglio et al.(1995)]{spinoglio95}
Spinoglio, L., Malkan, M. A., Rush, B., Carrasco, L., \& Recillas-Cruz, E. 1995, \apj, 453, 616
\bibitem[Stanford et al.(1990)]{stanford90}
Stanford, S. A., Sargent, A. I., Sanders, D. B., \& Scoville, N. Z.
1990, \apj, 349, 492
\bibitem[Taniguchi \& Ohyama(1998)]{taniguchi98}
Taniguchi, Y., \& Ohyama, Y. 1998, \apjl, 509, L89
\bibitem[Tielens \& Hollenbach(1985)]{tielens85}
Tielens, A. G. G. M., \& Hollenbach, D. 1985, \apj, 291, 747
\bibitem[Toomre \& Toomre(1972)]{toomre72}
Toomre, A., \& Toomre, J. 1972, \apj, 178, 623 (TT72)
\bibitem[Tully \& Pierce(2000)]{tully2000}
Tully, R. B., Pierce, M. J. 2000, \apj, 533, 744 
\bibitem[van der Hulst(1979)]{hulst79}
Hulst, J. M. van der 1979, \aap, 71, 131
\bibitem[van Driel, Gao, \& Monnier-Ragaigne(2001)]{driel01}
van Driel, W., Gao, Y., \&  Monnier-Ragaigne, D. 2001, \aap, in press
\bibitem[Welch et al.(1996)]{welch96}
Welch, W. J., et al. 1996, \pasp, 108, 93
\bibitem[Wynn-Williams et al.(1991)]{wynn91}
Wynn-Williams, C. G., Eales, S. A., Becklin, E. E., Hodaap, K.-W., Joseph, R. D.,
McLean, I. S., Simons, D. A., \& Wright, G. S. 1991, \apj, 377, 426
\bibitem[Young \& Knezek(1989)]{young89}
Young, J. S., \& Knezek, P. M. 1989, \apjl, 347, L55
\bibitem[Young \& Scoville(1991)]{young91}
Young, J. S., \& Scoville, N. Z. 1991, \araa, 29, 581
\bibitem[Yun, Ho, \& Lo(1994)]{yun94}
Yun, M. S., Ho, P. T. P., \& Lo, K. Y. 1994, \nat, 372, 530
\bibitem[Yun \& Scoville(1995)]{yun95}
Yun, M. S., \& Scoville, N. Z. 1995, \apjl, 451, L45
\bibitem[Zhou, Wynn-Williams, \& Sanders(1993)]{zhou93}
Zhou, S., Wynn-Williams, C. G., \& Sanders, D. B. 1993, \apj, 409, 149
\bibitem[Zwicky(1971)]{zwicky71}
Zwicky, F. 1971,  Catalogue of Selected Compact Galaxies and of  Post-Eruptive Galaxies
(Guemligen, Switzerland: F. Zwicky)

\end{thebibliography}
\end{document}